\documentclass[twocolumn]{aastex631}
\usepackage{multirow}
\usepackage{comment}

\newcommand{\ha}{${\rm H\alpha}$}

\shorttitle{VODKA: Spatially-resolved Optical Spectroscopy}
\shortauthors{Chen et al.}

\begin{document}

\title{Varstrometry for Off-nucleus and Dual sub-Kpc AGN (VODKA): Long-slit optical spectroscopic follow-up with Gemini/GMOS and HST/STIS}

\correspondingauthor{Yu-Ching Chen, Xin Liu}
\email{ycchen@jhu.edu; xinliuxl@illinois.edu}

\author[0000-0002-9932-1298]{Yu-Ching Chen}
\affiliation{Department of Physics and Astronomy, Johns Hopkins University, Baltimore, MD 21218, USA}
\affiliation{Department of Astronomy, University of Illinois at Urbana-Champaign, Urbana, IL 61801, USA}

\author[0000-0001-7681-9213]{Arran C. Gross}
\affiliation{Department of Astronomy, University of Illinois at Urbana-Champaign, Urbana, IL 61801, USA}

\author[0000-0003-0049-5210]{Xin Liu}
\affiliation{Department of Astronomy, University of Illinois at Urbana-Champaign, Urbana, IL 61801, USA}
\affiliation{National Center for Supercomputing Applications, University of Illinois at Urbana-Champaign, Urbana, IL 61801, USA}

\author[0000-0003-1659-7035]{Yue Shen}
\affiliation{Department of Astronomy, University of Illinois at Urbana-Champaign, Urbana, IL 61801, USA}
\affiliation{National Center for Supercomputing Applications, University of Illinois at Urbana-Champaign, Urbana, IL 61801, USA}

\author[0000-0001-6100-6869]{Nadia L. Zakamska}
\affiliation{Department of Physics and Astronomy, Johns Hopkins University, Baltimore, MD 21218, USA}

\author[0000-0003-4250-4437]{Hsiang-Chih Hwang}
\affiliation{School of Natural Sciences, Institute for Advanced Study, Princeton, 1 Einstein Drive, NJ 08540, USA}

\author[0000-0001-5105-2837]{Ming-Yang Zhuang}
\affiliation{Department of Astronomy, University of Illinois at Urbana-Champaign, Urbana, IL 61801, USA}



\begin{abstract}

High spatial and spectral resolution observations are essential for identifying sub-arcsecond dual and lensed quasars and confirming their redshifts. We present Gemini/GMOS and HST/STIS optical spectra for 27 dual quasar candidates selected based on their variability-induced astrometric noise or double detections in Gaia (the VODKA project). From this follow-up, we spectroscopically identify 11 star superpositions and 7 dual/lensed quasars. Among the remaining targets, 2 are likely dual/lensed quasars based on additional radio imaging, while the rest are quasars with unknown companions. 
Without prior photometric or spectroscopic selection, we find the star contamination rate to be 41-67\%, while the dual/lensed quasar fraction is $\gtrsim$ 26\% in the follow-up VODKA sample. However, when combined with existing unresolved spectra and spatially-resolved two-band color cuts, the dual/lensed quasar fraction can be increased to $\gtrsim$ 67\%. 
Our study highlights the need for high-quality spectral data, including a signal-to-noise ratio of at least 20, spatial resolution that is at least twice finer than the source separation, and a spectral resolution of $R \gtrsim 1000$, in order to separate close sources, exclude stellar superpositions, and reliably identify dual quasars.

\end{abstract}



\section{Introduction} \label{sec:intro}

Most nearby massive galaxies harbor a supermassive black hole (SMBH) at their centers \citep{Magorrian1998, KormendyHo2013}. Cold dark matter models often depict hierarchical structure formation and galaxy mergers \citep{White1991, Navarro1996, Cole2000}. 
When two galaxies merge, gas can accrete onto the central SMBHs as tidal forces generate non-axisymmetric structures in the companion galaxy, which torque the gas and drive it toward the nuclei \citep{Hernquist1989,Barnes1996,Mihos1996,Hopkins2010}. In addition to these tidal torque-induced inflows, ram-pressure sweeping at pericentric passage can effectively decouple gas from stars, reducing the gas's angular momentum and further enhancing central inflows \citep{Barnes2002,Capelo2017}.
If both SMBHs are active and still within their host galaxies' potential, they are known as dual quasars. While cosmological simulations show increased quasar activity and dual quasar fraction in the late-stage mergers \citep{Steinborn2016,DeRosa2019,Rosas-Guevara2019,Volonteri2022,ChenN2023}, the observational evidence for the link between quasar activity and mergers is still debated \citep{Mechtley2016,Fan2016,Ellison2019,Breiding2024}. The dual quasar phase is an excellent stage to study quasar triggering and its connection to galaxy mergers \citep{Volonteri2022}. 

Identifying dual quasars is challenging due to their rarity and close separations, particularly at higher redshifts ($z>1$) when the angular diameter distance is large (1\arcsec $\sim$ 8 kpc at $z\sim2$). In recent years, large sky surveys with high angular resolution, such as Gaia, have become valuable tools for finding dual and lensed quasars \citep{Lemon18,Lemon19,ChenYC2022,Lemon23}. Despite the introduction of various techniques to discover small-separation ($\lesssim$1") dual quasar candidates \citep{Mannucci2022, HwangShen2020}, only a small fraction of these candidates have been confirmed \citep{ChenYC2023a,Mannucci2023, Ciurlo2023,Scialpi2024}. One notable effort in finding close-separation dual quasars is the Varstrometry for Off-nucleus and Dual sub-Kpc AGN (VODKA) project \citep{HwangShen2020, ShenHwang2019}. The VODKA project aims to discover unresolved dual quasars by analyzing centroid jitters from the non-coherent quasar light from two sources. With dozens of dual quasar candidates discovered by the VODKA project, spectroscopic follow-ups are essential to confirm their nature \citep{ChenYC2022}.

In this paper, we present optical spectroscopic observations for 27 targets from the VODKA project. We provide classifications based on spectra from the Hubble Space Telescope (HST) and the Gemini Observatory, along with other auxiliary data. 
The paper is organized as follows. Section \ref{sec:data} describes the data reduction and analysis procedures, as well as the spectral fitting details. In section \ref{sec:results}, we show the HST and Gemini spectra for the 27 targets and provide our best classifications. Section \ref{sec:discussion} presents the sample category breakdown, highlights the cautionary case of J1649+0812 with an apparent velocity offset, and discusses implications for future spectroscopic studies. We summarize our findings in section \ref{sec:conclusion}. We adopt a flat $\Lambda$CDM cosmology with $\Omega_\Lambda=0.7$, $\Omega_m=0.3$, and $H_0=70\,{\rm km\,s^{-1}Mpc^{-1}}$ throughout this paper. 

\section{Observations, Data reduction and analysis} \label{sec:data}

\subsection{Target selection: dual quasar candidates}
The 27 targets in this paper are the dual/lensed quasar candidates with small separations ($<$1") discovered using Gaia astrometry and HST images \citep{ChenYC2022}. The sample consists of 14 Gaia-resolved targets with separations $\gtrsim$0\farcs5 and 13 Gaia-unresolved targets selected by the astrometric technique, Varstrometry. The Varstrometry technique utilizes quasar's strong variability and the non-synchronized emission of two nuclei to discover unresolved dual quasars \citep{HwangShen2020,ShenHwang2019}. To confirm the nature of these candidates, we conducted a series of spectroscopic follow-up observations using HST and the Gemini Observatory. In addition to the spectroscopic campaign, we carried out imaging programs with the Very Large Array in radio and HST in the near-infrared to detect lens galaxies, study tidal features, and investigate radio properties. These findings are reported in a companion paper \citep{Gross2024}.

\subsection{HST/STIS} \label{sec:data-hst}

Out of 27 targets, 22 were observed with HST using the Space Telescope Imaging Spectrograph (STIS) during cycles 28 and 29 in 2021-2022 under programs GO-16210 and GO-16887 (PI: X. Liu). All the HST data used in this paper can be found in MAST: \dataset[10.17909/a3ac-ny86]{http://dx.doi.org/10.17909/a3ac-ny86}.
The observations utilized the G750L grating with a 52\arcsec$\times$0.2\arcsec aperture, covering observed wavelengths from 5240\AA\ to 10270\AA, with a resolving power of R $\sim$500. The slit direction was rotated to capture both nuclei, guided by the HST-resolved images \citep{ChenYC2022}. Exposure times and additional observational details can be found in \autoref{tab:obs}.

The STIS spectra were reduced using the \texttt{stis\_cti} package, which corrects for trails and artifacts caused by charge transfer inefficiency effects \citep{Anderson2010}. Following the standard STIS calibration pipeline, \texttt{calstis}, we applied flat-field correction, bad-pixel removal, and cosmic ray rejection. The final one-dimensional (1D) spectra for each nucleus were extracted using a boxcar aperture of 7 pixels (0\farcs35). \autoref{fig:spectra_stis1} and \autoref{fig:spectra_stis2} present the calibrated 1D spectra of all the targets observed with HST/STIS.

\subsection{Gemini/GMOS} \label{sec:data-gemini}
16 targets were observed using the Gemini Multi-Object Spectrographs (GMOS) on both the Gemini-North and Gemini-South telescopes in 2021 and 2022, under several programs (GN-2020A-DD-106, GS-2020A-DD-106, GN-2022A-Q-139, GS-2022A-Q-148; PI: X. Liu, and GN-2020A-Q-232; PI: Y.-C. Chen).
The observations used the R150 grating centered around $\sim$700 nm with slit widths ranging from 0\farcs5 to 0\farcs75, covering the observed wavelength range of 4000\AA\ to 10040\AA, with a resolving power $R$ of $\sim$420–630.
The typical seeing conditions were $\lesssim$0\farcs75 (within the 70th percentile in optical), and for targets with smaller separations ($<$0\farcs5), seeing was $\lesssim$0\farcs4 (within the 20th percentile in optical).
Although photometric cloud conditions were not required, most targets were observed under photometric conditions, with no significant signal loss due to clouds. As with the HST/STIS spectra, the slit direction was rotated to capture both nuclei. Standard stars EG131, LTT1788, Wolf1346, and LTT4816 were observed with a similar setup for flux calibration. The exposure times and additional observational details are provided in \autoref{tab:obs}. 

We reduced the raw exposures using the Python wrapper of the \texttt{iraf} software.  We perform bias subtraction and flat-field correction. Wavelength calibration was completed using arc exposures taken with the same configuration, and flux calibration was based on standard star exposures with the same grating. We dithered the exposures along both the spatial and spectral directions. The final 2D exposures were combined to eliminate artifacts, remove bad pixels, and fill detector gaps.

The typical seeing conditions for the Gemini observations ranged from 0\farcs5 to 0\farcs8, comparable to the nuclear separations of our targets. Since many of the targets were only marginally resolved, we opted for a two-component Gaussian fit rather than boxcar extraction. At each spectral element, we fit two Gaussian components with fixed separations along the spatial direction, allowing the centroids and FWHMs to vary initially to account for distortions and seeing variations across the spectral direction. The centroids and FWHMs were then refitted with a cubic spline function. Final flux densities were derived by fitting two Gaussian components with fixed separations and wavelength-dependent centroids and FWHMs, as determined by the cubic spline fit. \autoref{fig:spectra_gmos} presents the calibrated 1D spectra of all the targets observed with Gemini/GMOS.

\begin{deluxetable*}{cccccccc}
 \tablecaption{Observation details for the 27 targets from Gemini and HST.
 \label{tab:obs}}
 \tablehead{ \colhead{Abbreviated Name} & \colhead{Obs. Date} & \colhead{Exp. Time} & \colhead{SNR} & \colhead{$R$} & \colhead{Telescope/Instrument} & \colhead{Grating} & \colhead{Program ID} \\ 
 \colhead{(J2000)} &  \colhead{(UT)} &\colhead{(min)} & & &  & & } 
 \colnumbers
 \startdata
WISE J0241+7801 & 2022-05-11 & 40 & 5,1 & \multirow{14}{*}{500} & \multirow{14}{*}{HST/STIS} & \multirow{14}{*}{G750L} & \multirow{14}{*}{HST-GO-16210} \\
WISE J0246+6922 & 2021-12-28 & 39 & 9,6 & & & & \\
WISE J0536+5038 & 2021-12-12 & 30 & 12,5 & & & &\\
SDSS J0748+3146 & 2022-01-03 & 33 & 2,2 & & & &\\
SDSS J0749+2255 & 2021-02-16 & 36 & 5,2 & & & &\\
SDSS J0753+4247 & 2022-01-17 & 31 & 6,2 & & & &\\
SDSS J0823+2418 & 2022-01-13 & 30 & 7,5 & & & &\\
SDSS J0841+4825 & 2021-02-10 & 32 & 3,3 & & & &\\
SDSS J0904+3332 & 2021-02-28 & 36 & 5,4 & & & &\\
WISE J1613$-$2644& 2021-02-27 & 36 & 2,1 & & & &\\
WISE J1755+4229 & 2022-04-01 & 35 & 12,2 & & & &\\
WISE J1852+4833 & 2021-05-13 & 36 & 6,2 & & & &\\
WISE J2048+6258 & 2021-02-15 & 39 & 4,2 & & & &\\
SDSS J2122$-$0026& 2021-08-03& 30 & 4,3 & & & &\\
\hline
WISE J0348$-$4015 & 2022-07-09& 35 & 2,1 & \multirow{8}{*}{500} &\multirow{8}{*}{HST/STIS} & \multirow{8}{*}{G750L} & \multirow{8}{*}{HST-GO-16887} \\
SDSS J1327+1036 & 2022-05-28 & 31 & 2,1 & & & &\\
SDSS J1648+4155 & 2022-03-28 & 35 & 8,2 & & & &\\
WISE J1649+0812 & 2022-04-20 & 33 & 7,3 & & & &\\
WISE J1711$-$1611 & 2022-06-12 & 32 & 3,2 & & & &\\
WISE J1937$-$1821 & 2022-06-11 & 34 & 2,2 & & & &\\
WISE J2050$-$2947 & 2022-07-20 & 31 & 6,3 & & & &\\
WISE J2218$-$3322 & 2022-09-05 & 34 & 1,0.3 & & & &\\
\hline
SDSS J0841+4825 & 2020-05-21 & 31 & 71,51& \multirow{2}{*}{420}  &  \multirow{2}{*}{Gemini/GMOS} & \multirow{2}{*}{R150} & \multirow{2}{*}{GN-2020A-DD-106} \\
SDSS J0904+3332 & 2020-05-21 & 31 & 54,15 &  & & & \\
\hline
WISE J1314$-$4912 & 2020-02-23 & 31 & 43,33 & 420 & Gemini/GMOS & R150 & GS-2020A-DD-106 \\
\hline
WISE J1732$-$1335 & 2020-08-13 & 40 & 42,34 & \multirow{4}{*}{420} & \multirow{4}{*}{Gemini/GMOS} &  \multirow{4}{*}{R150} & \multirow{4}{*}{GN-2020A-Q-232} \\
WISE J1804+3230 & 2020-08-14 & 40 & 38,35 &  & & &\\
WISE J1852+4833 & 2020-07-02 & 60 & 52,22 &  & & &\\
WISE J1857+7048 & 2020-07-29 & 60 & 40,15 &  & & &\\
\hline
SDSS J0749+2255 & 2022-04-20 & 31 & 29,19 & 630 &  \multirow{5}{*}{Gemini/GMOS} & \multirow{5}{*}{R150} & \multirow{5}{*}{GN-2022A-Q-139} \\
SDSS J0823+2418 & 2022-04-20 & 17 & 40,36 & 420 & & &\\
SDSS J1225+4831 & 2022-04-20 & 23 & 31,24 & 420 & & &\\
SDSS J1648+4155 & 2022-04-21 & 53 & 49,14 & 630 & & &\\
WISE J1649+0812 & 2022-04-21 & 23 & 42,37 & 630 & & &\\
\hline
WISE J1711$-$1611 & 2022-04-01 & 23 & 23,18 & 420 & \multirow{4}{*}{Gemini/GMOS} & \multirow{4}{*}{R150} & \multirow{4}{*}{GS-2022A-Q-148} \\
WISE J1937$-$1821 & 2022-06-11 & 26 & 21,18 & 420 & & & \\
WISE J2050$-$2947 & 2022-06-12 & 20 & 33,21 & 420 & & & \\
WISE J2218$-$3322 & 2022-11-17 & 40 & 16,10 & 630 & & & \\
    \enddata
 \tablecomments{Column 1: Target name. Column 2: Observation date. Column 3: Exposure time. Column 4: Median signal-to-noise ratio per spectral element for each source. Column 5: Spectral resolution $R=\lambda/\Delta\lambda$. Column 6: Telescope and instrument. Column 7: Grating name. Column 8: Program ID}
\end{deluxetable*}



\section{Results} \label{sec:results}

\subsection{Spatially-resolved spectra}

\begin{figure*}
  \centering
    \includegraphics[width=0.325\textwidth]{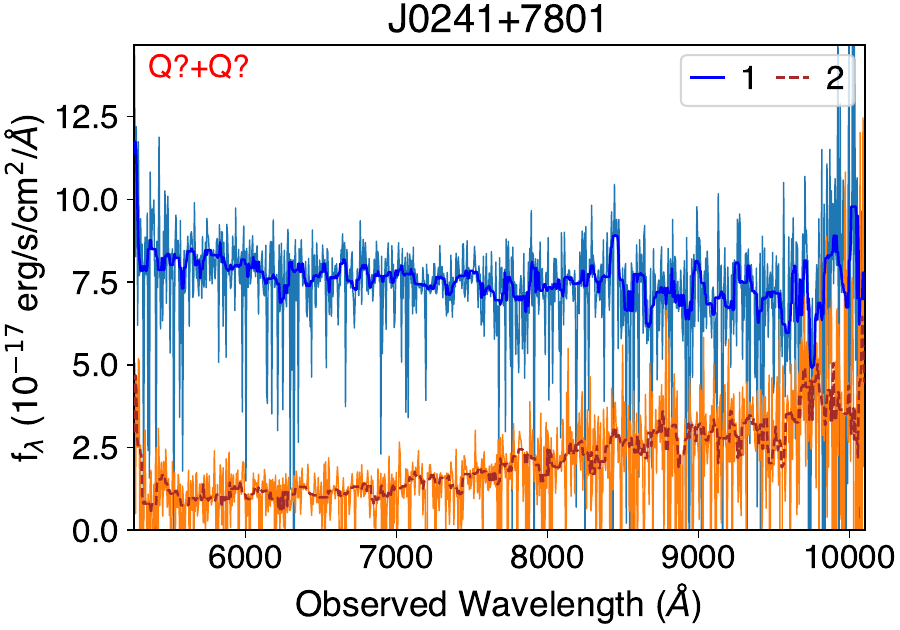}
    \includegraphics[width=0.325\textwidth]{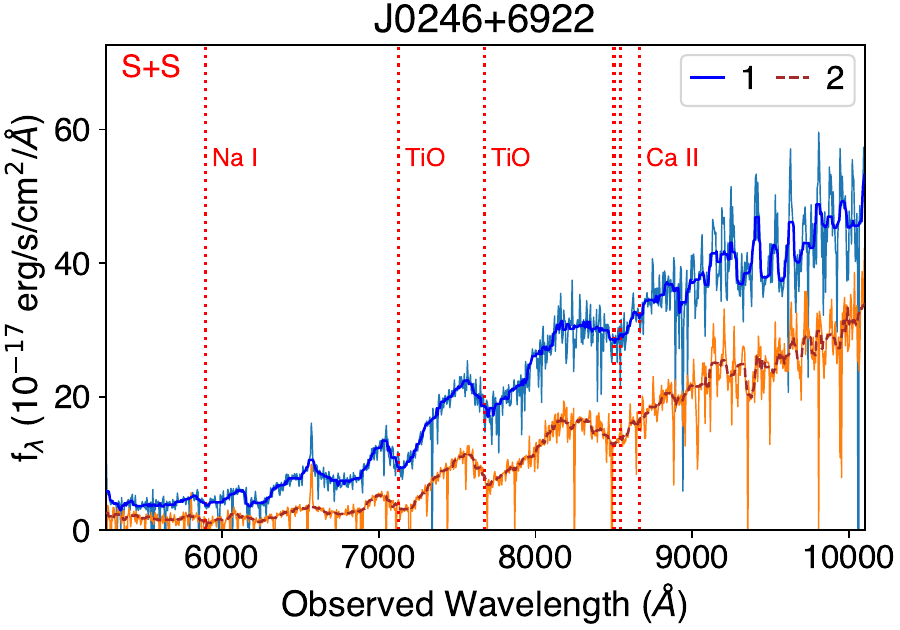}
    \includegraphics[width=0.325\textwidth]{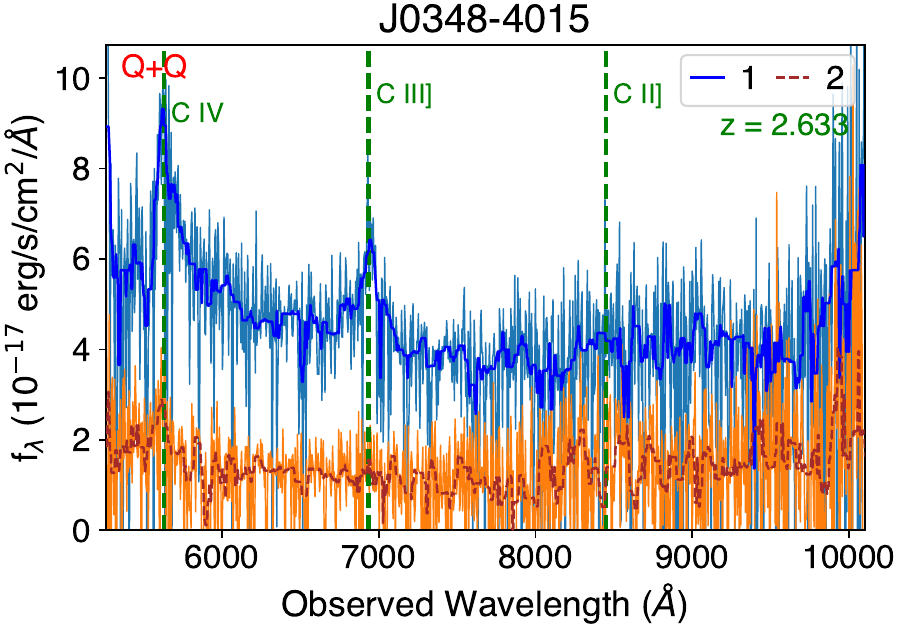}
    \includegraphics[width=0.325\textwidth]{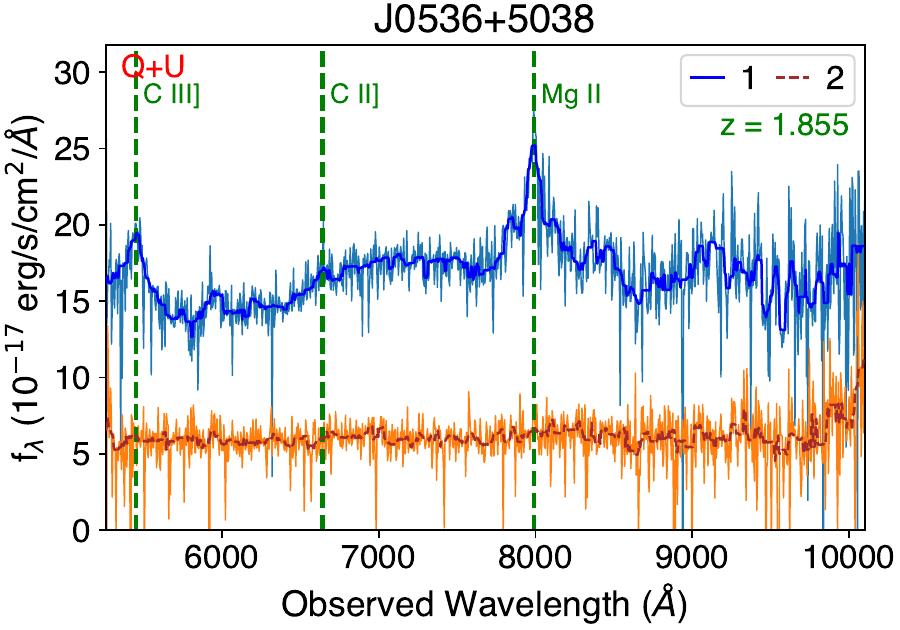}
    \includegraphics[width=0.325\textwidth]{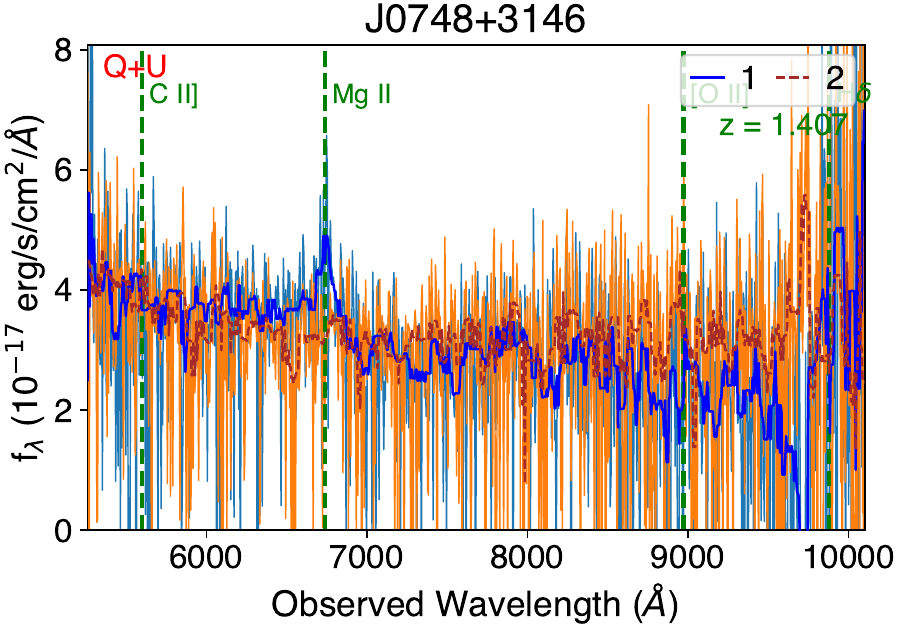}
    \includegraphics[width=0.325\textwidth]{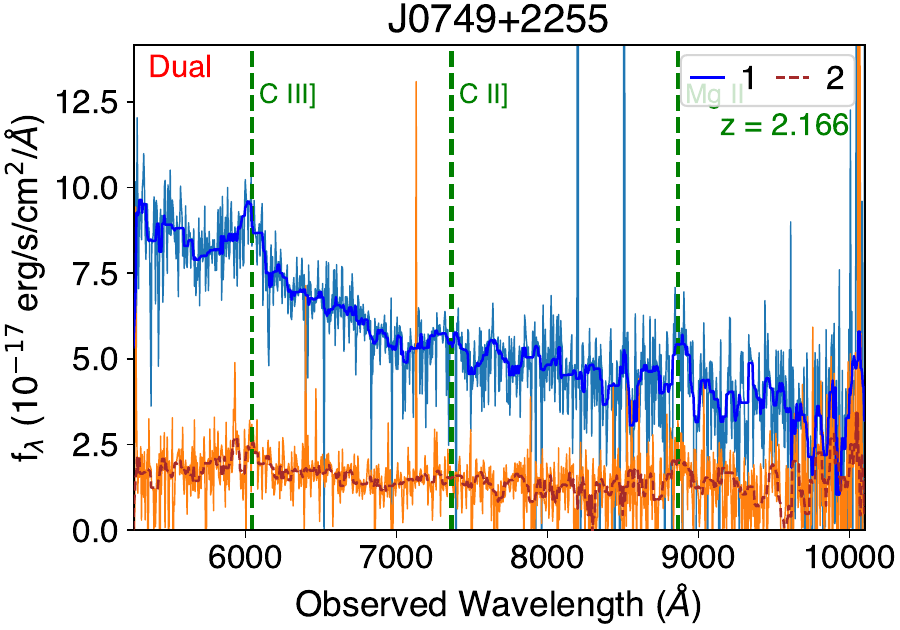}
    \includegraphics[width=0.325\textwidth]{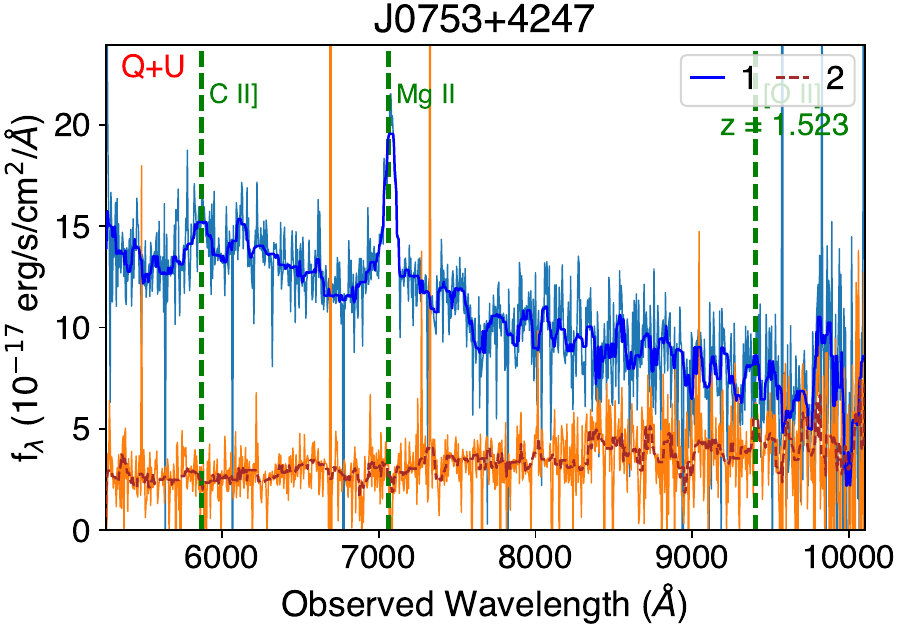}
    \includegraphics[width=0.325\textwidth]{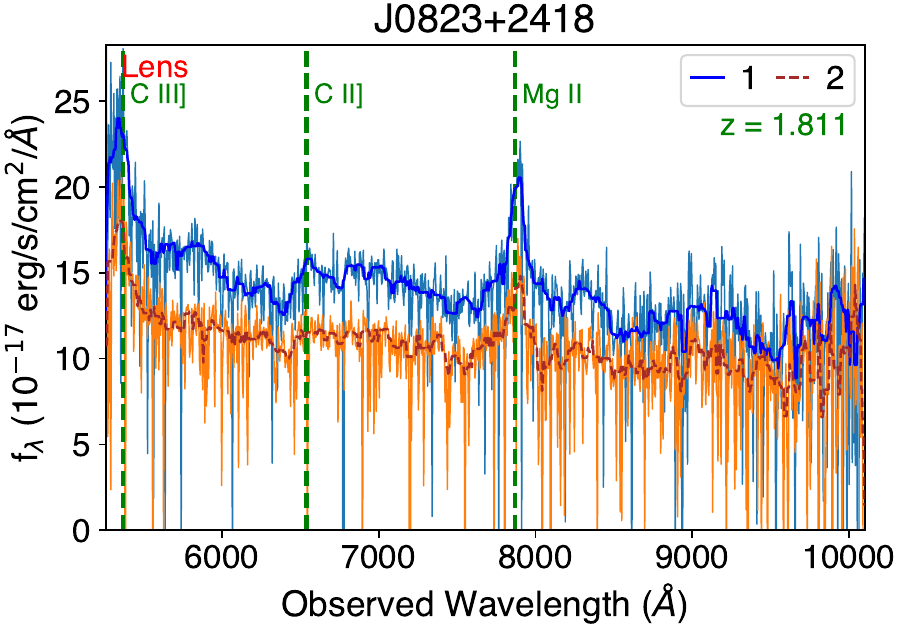}
    \includegraphics[width=0.325\textwidth]{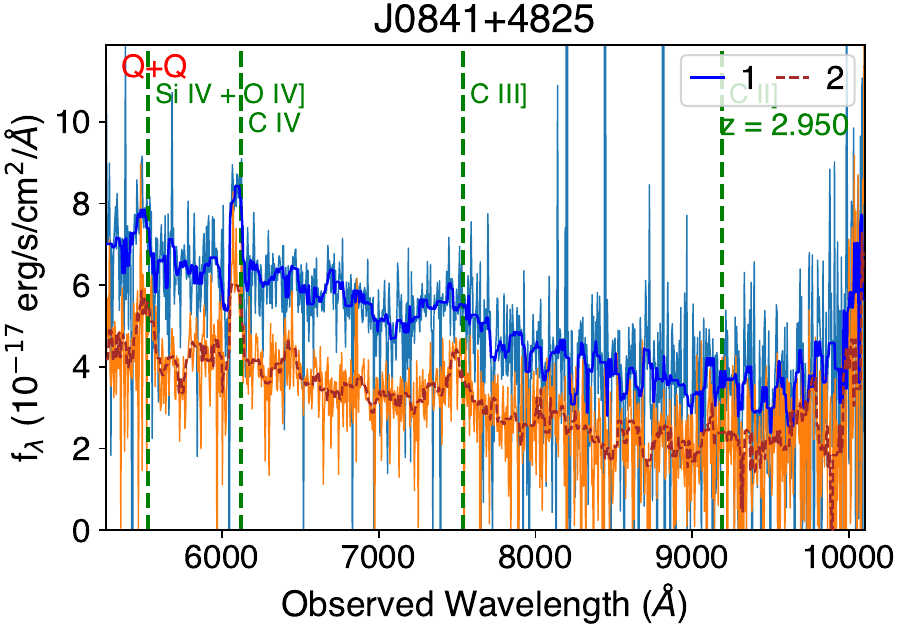}
    \includegraphics[width=0.325\textwidth]{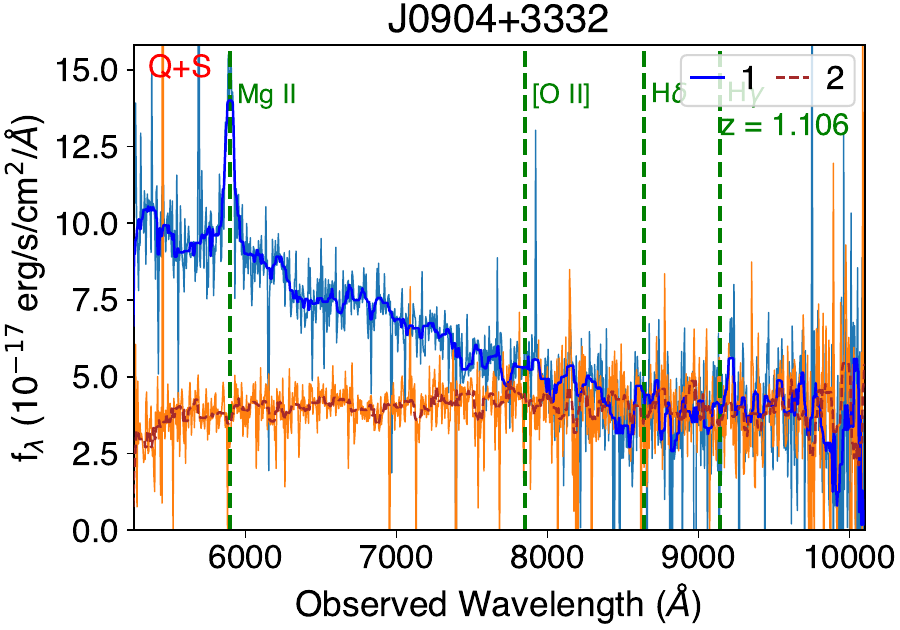}
    \includegraphics[width=0.325\textwidth]{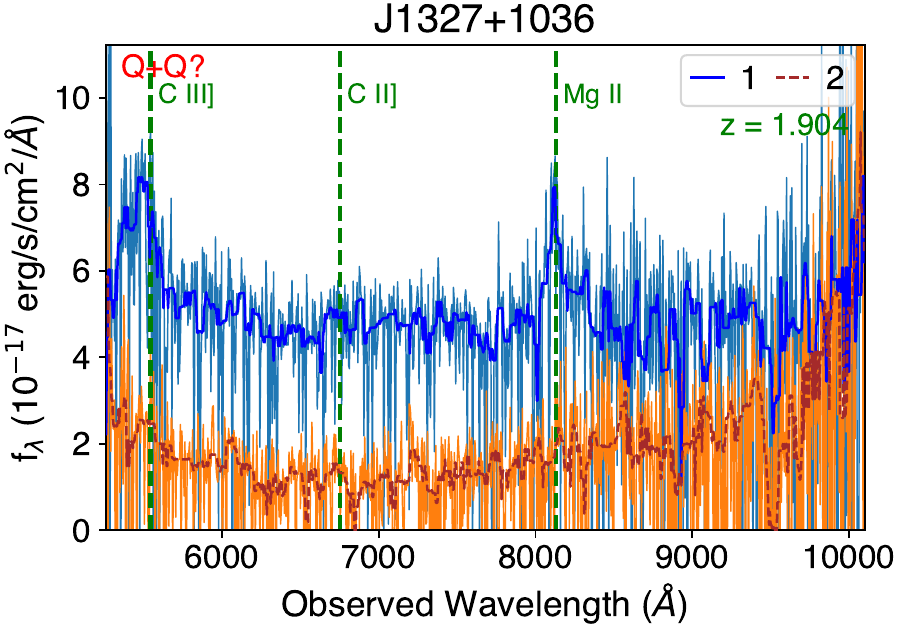}
    \includegraphics[width=0.325\textwidth]{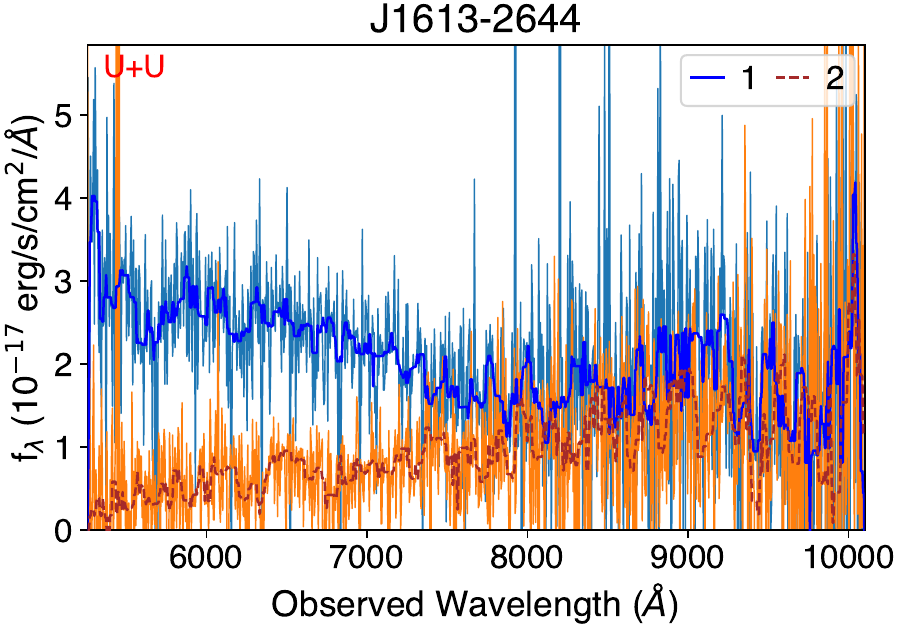}
    \includegraphics[width=0.325\textwidth]{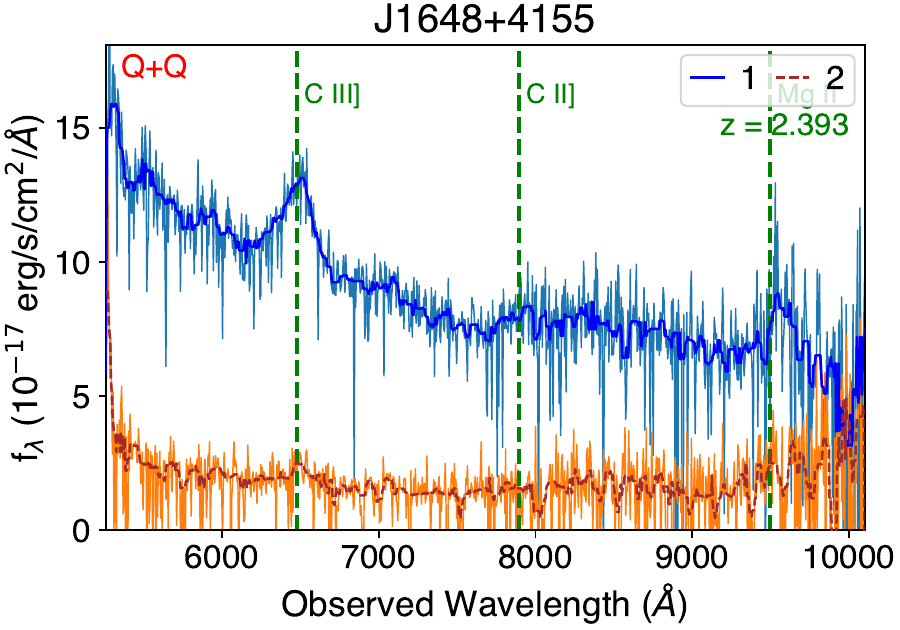}
    \includegraphics[width=0.325\textwidth]{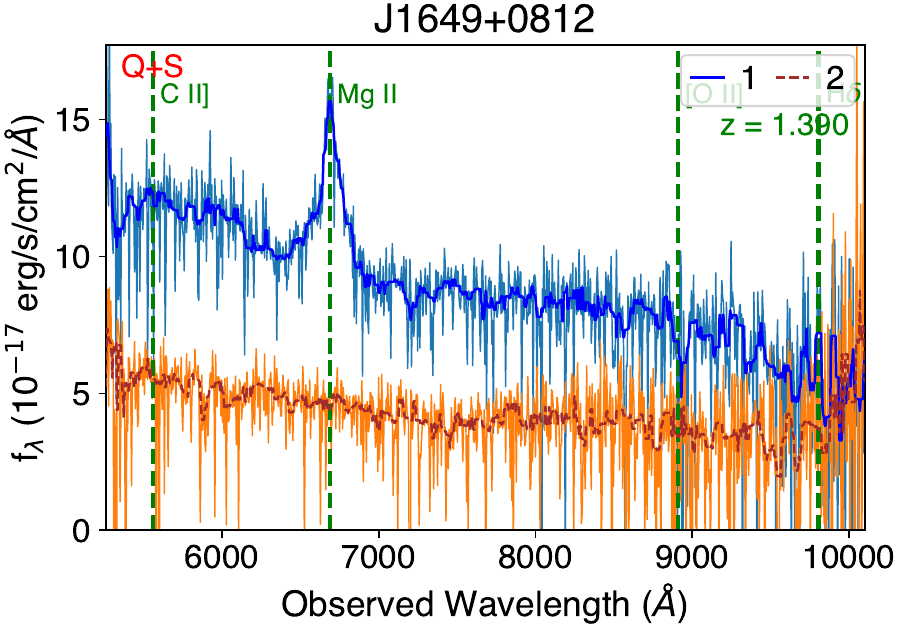}
    \includegraphics[width=0.325\textwidth]{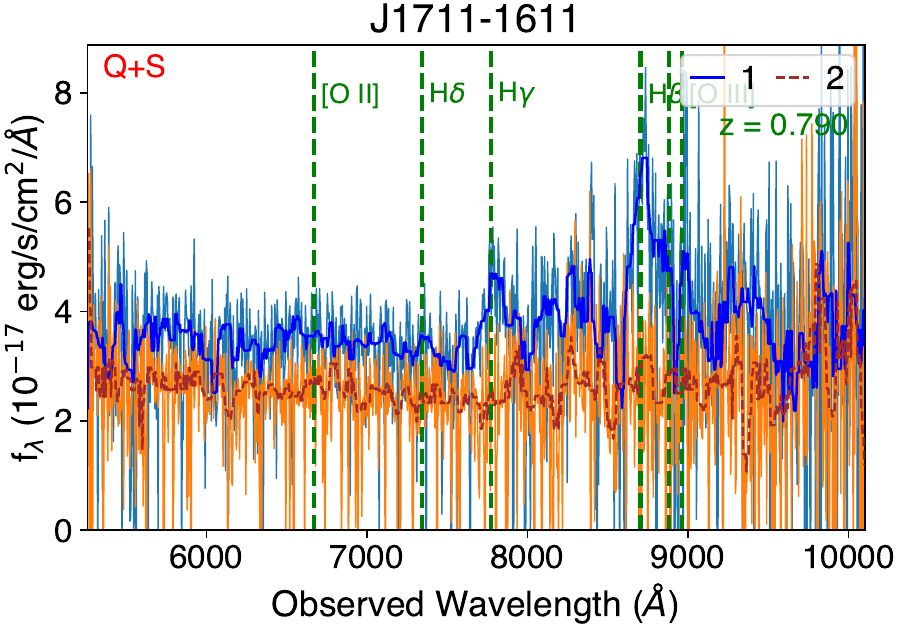}
    
    \caption{HST STIS optical spectra of 22 targets (15 in this figure and 7 in \autoref{fig:spectra_stis2}). Spatially resolved spectra for both sources are shown in solid blue and dashed orange lines. Green vertical dashed lines indicate the quasar's emission lines at the systemic redshift, displayed in the top-right corner. Red vertical dotted lines indicate stellar absorption features for star superposition. Final classifications based on both Gemini and/or HST spectra are shown in the top-left corner. (Q: quasar, S: star, U: unknown, Dual: dual quasar, Lens: lensed quasar, ?: likely).
    }
    \label{fig:spectra_stis1}
\end{figure*}

\begin{figure*}
  \centering
    \includegraphics[width=0.325\textwidth]{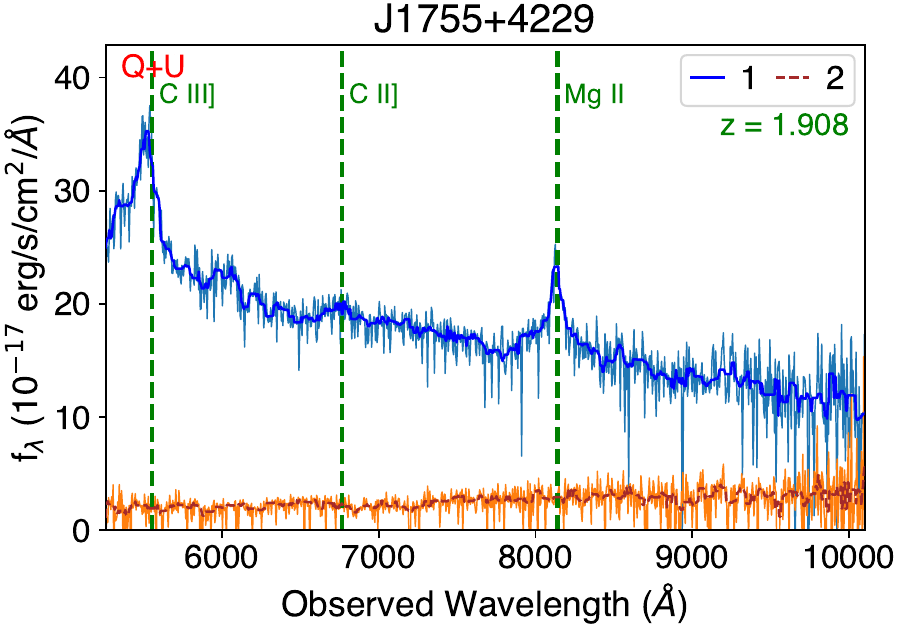}
    \includegraphics[width=0.325\textwidth]{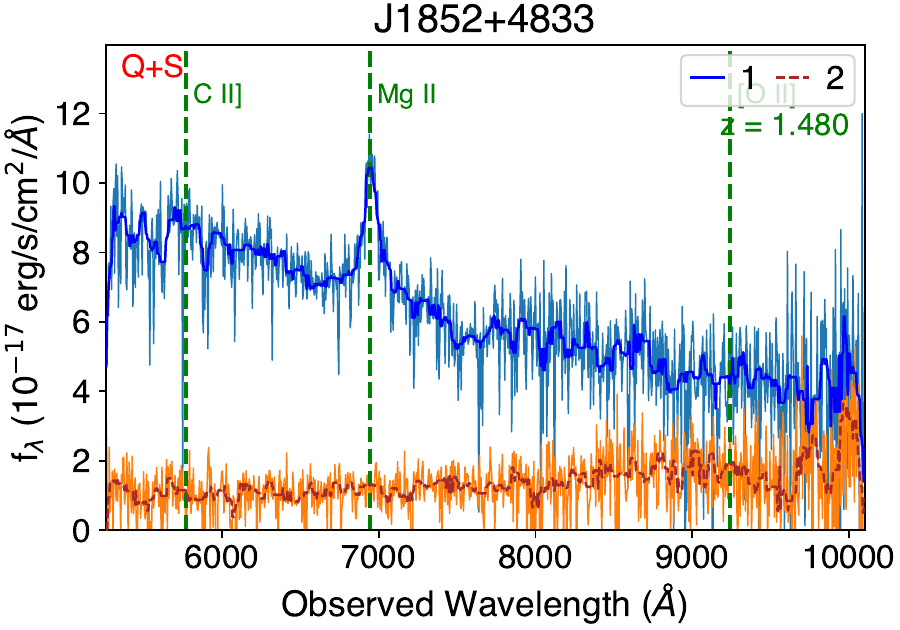}
    \includegraphics[width=0.325\textwidth]{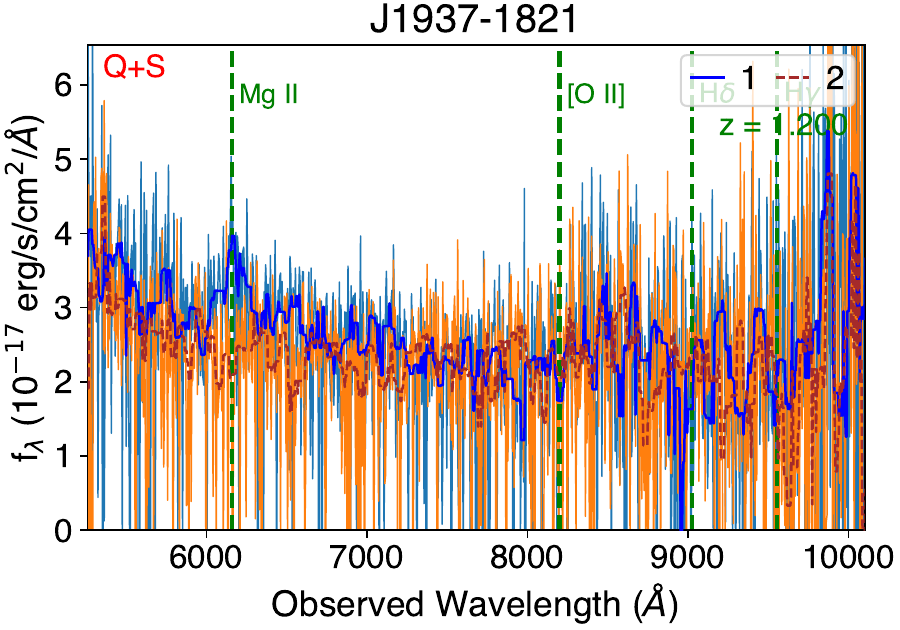}
    \includegraphics[width=0.325\textwidth]{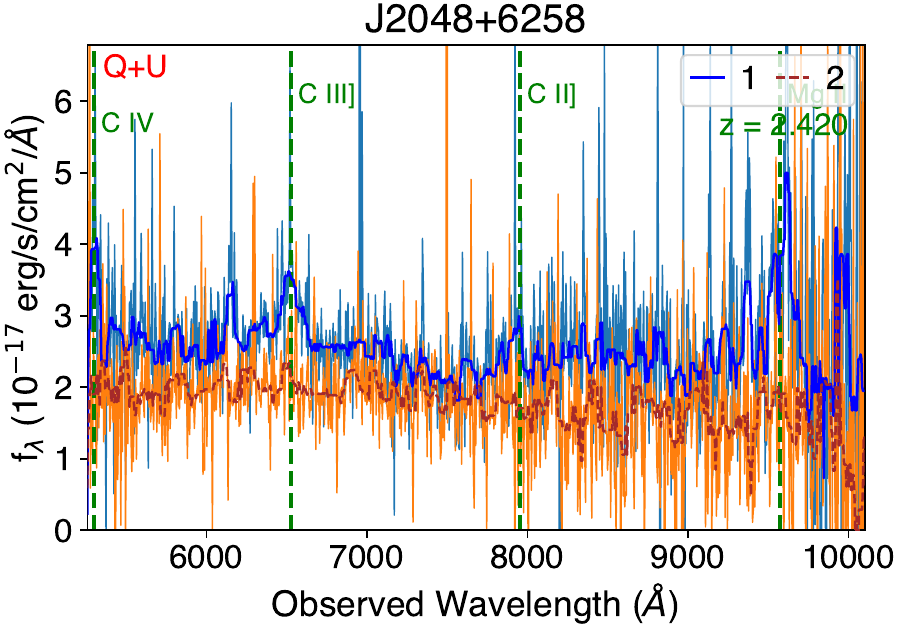}
    \includegraphics[width=0.325\textwidth]{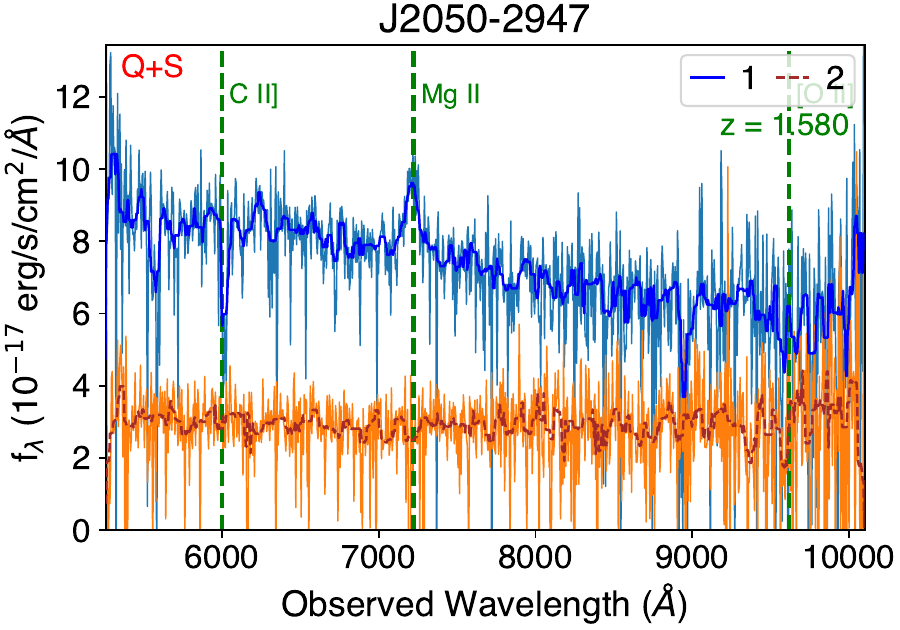}
    \includegraphics[width=0.325\textwidth]{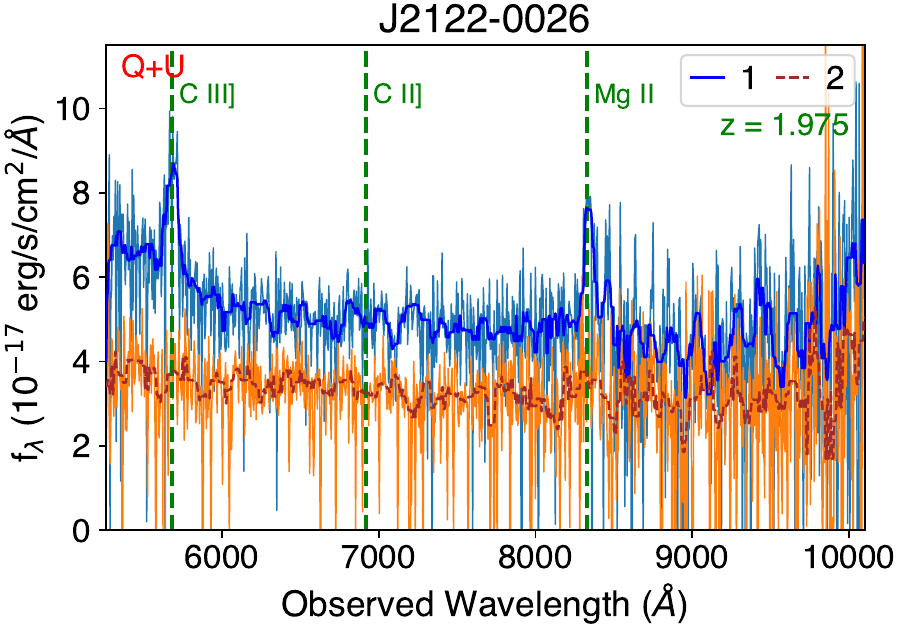}
    \includegraphics[width=0.325\textwidth]{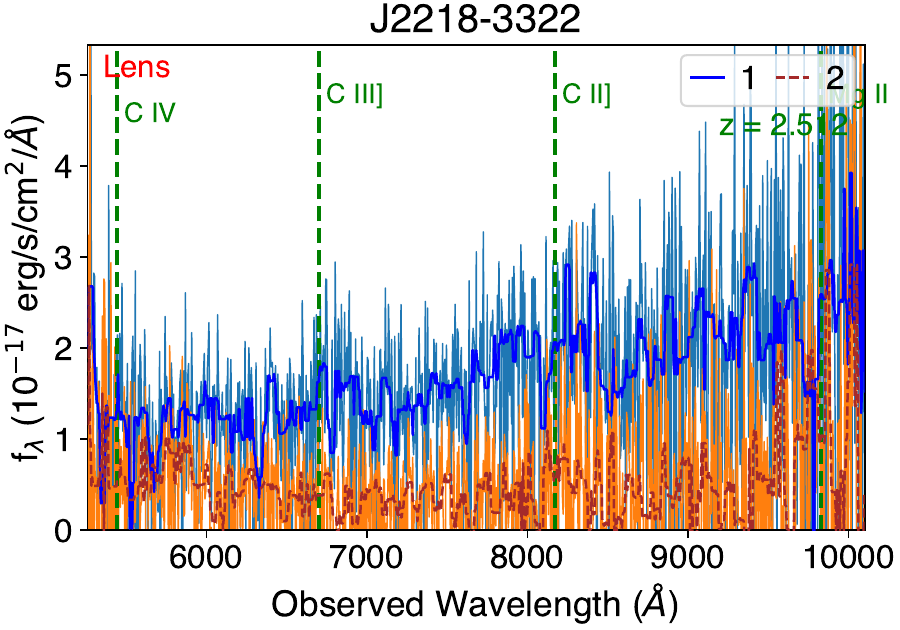}
    
    \caption{HST STIS optical spectra of 22 targets (7 in this figure and 7 in \autoref{fig:spectra_stis1}). Notations are the same as those in \autoref{fig:spectra_stis1}
    }
    \label{fig:spectra_stis2}
\end{figure*}

We present the optical spectra of the 27 targets obtained from Gemini/GMOS and HST/STIS in this section. \autoref{fig:spectra_stis1} and \autoref{fig:spectra_stis2} show the HST/STIS spectra of 22 targets. All targets are spatially resolved in the HST/STIS spectra, benefiting from HST’s high angular resolution ($\sim$0\farcs075 at 800 nm). For 80\% of the targets, we identify the quasars' broad emission lines in at least one of the sources. However, given the faintness of our targets ($f_\lambda$ $\lesssim$ 10$^{-16}$ erg s$^{-1}$ cm$^{-2}$ \AA$^{-1}$), the typical signal-to-noise ratios are $\lesssim$ 10, making it challenging to identify stellar absorption lines.

\begin{figure*}
  \centering
    \includegraphics[width=0.32\textwidth]{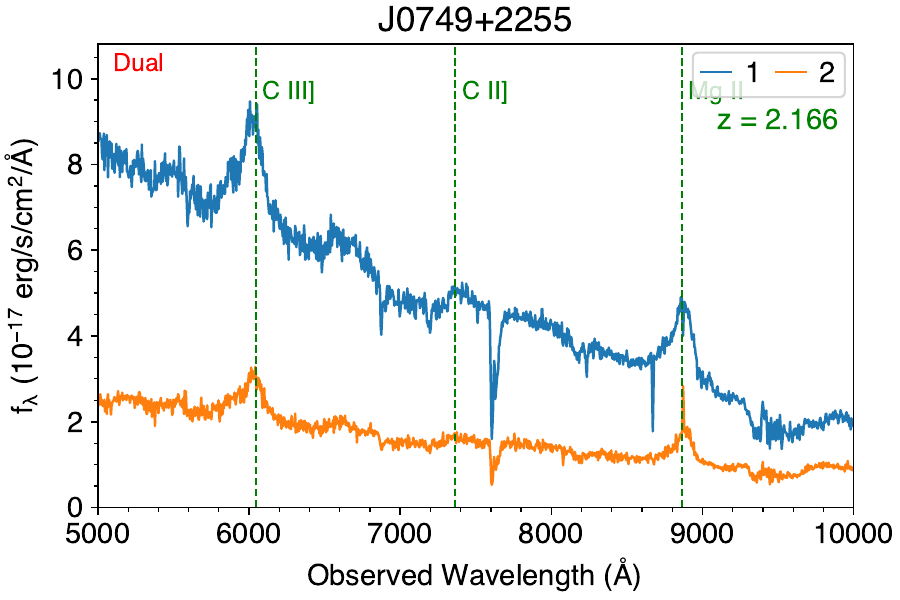}
    \includegraphics[width=0.32\textwidth]{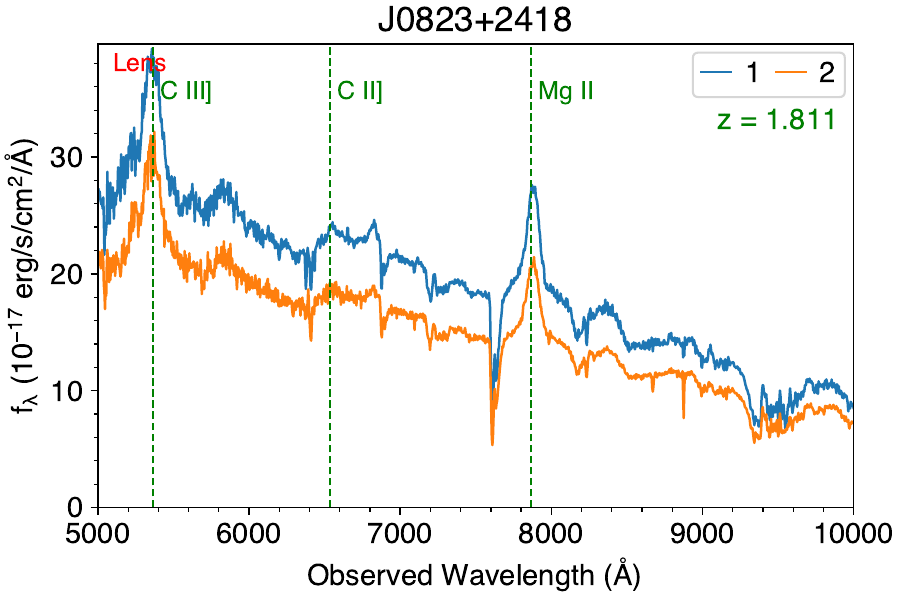}
    \includegraphics[width=0.32\textwidth]{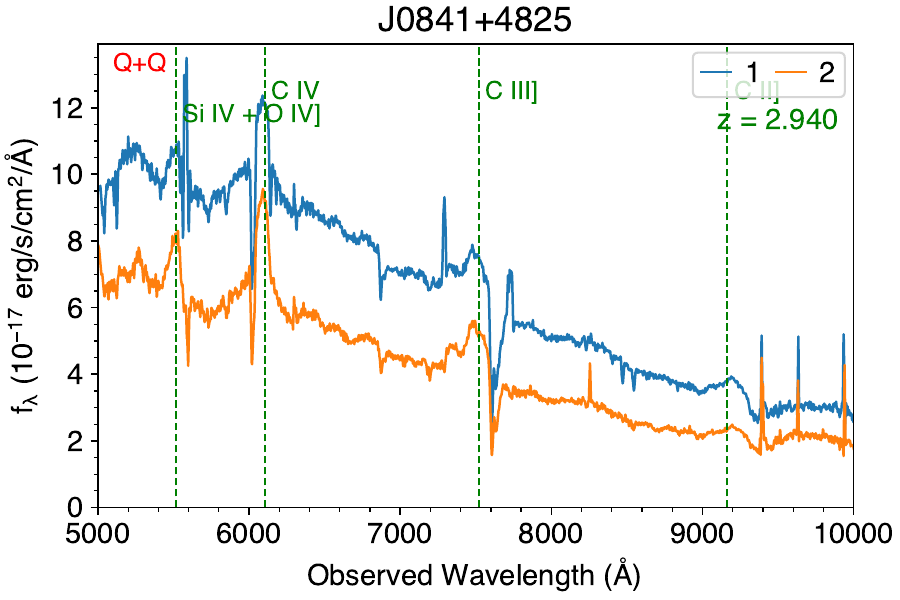}
    \includegraphics[width=0.32\textwidth]{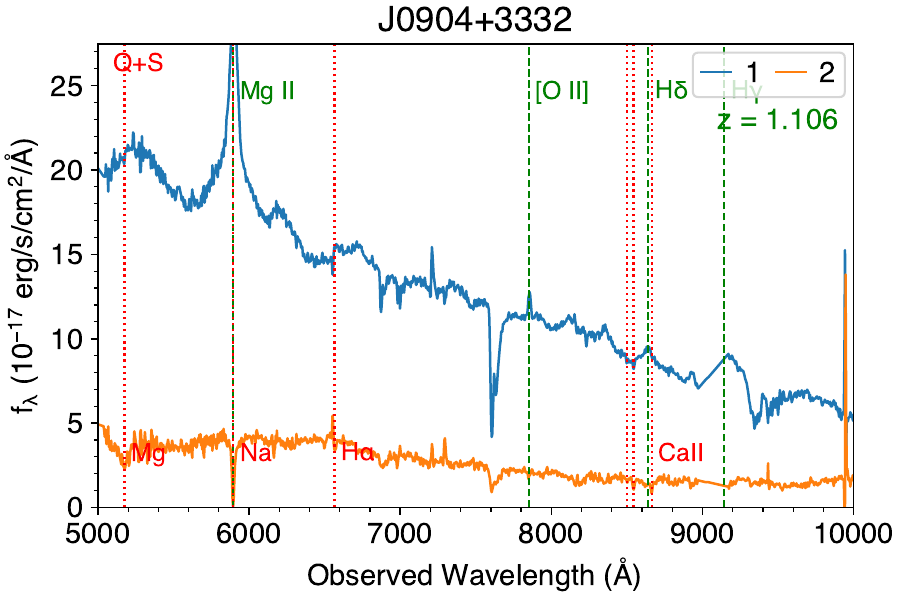}
    \includegraphics[width=0.32\textwidth]{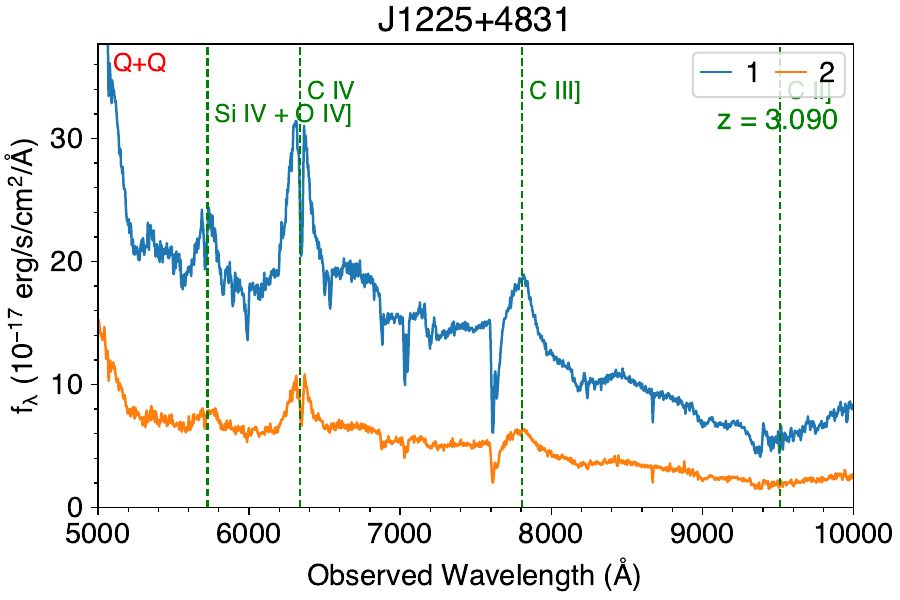}
    \includegraphics[width=0.32\textwidth]{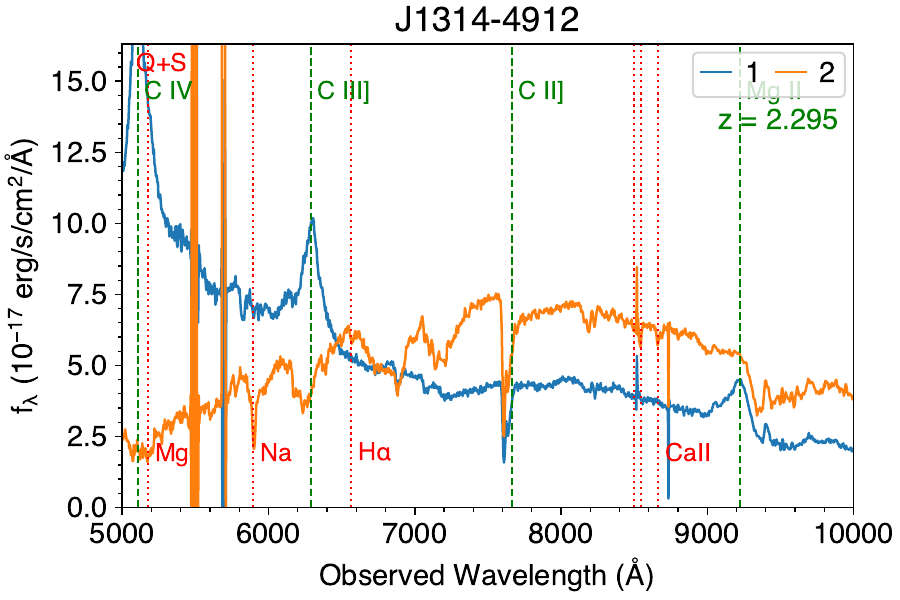}
    \includegraphics[width=0.32\textwidth]{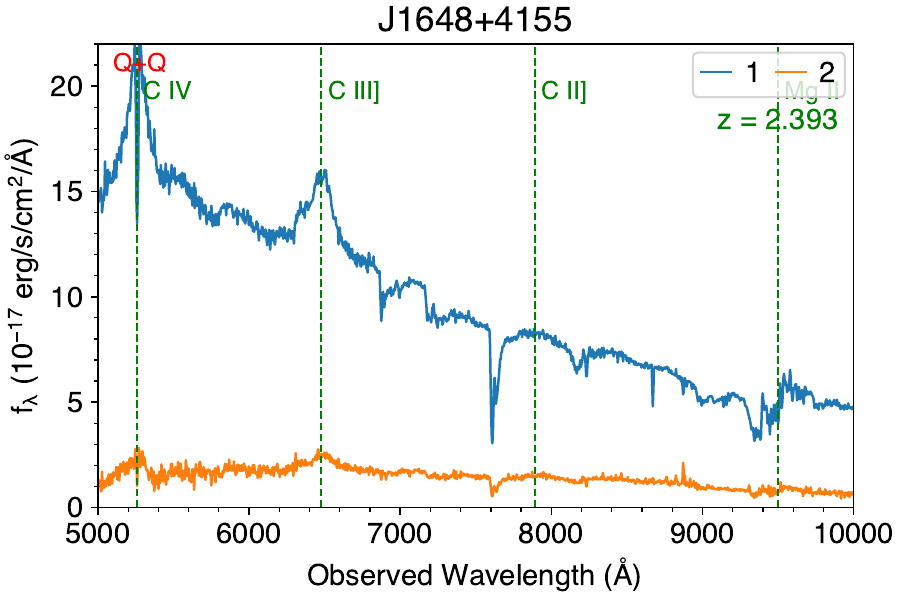}
    \includegraphics[width=0.32\textwidth]{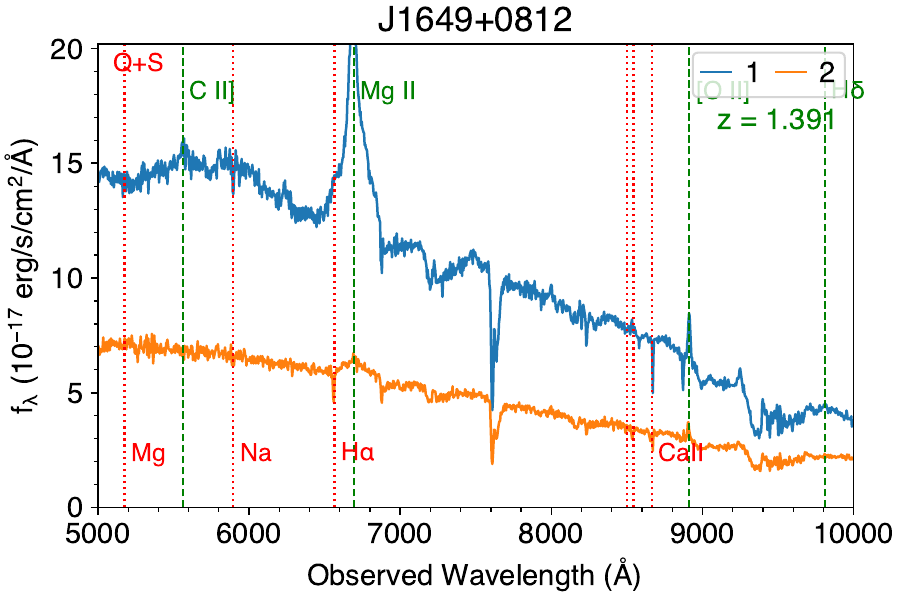}
    \includegraphics[width=0.32\textwidth]{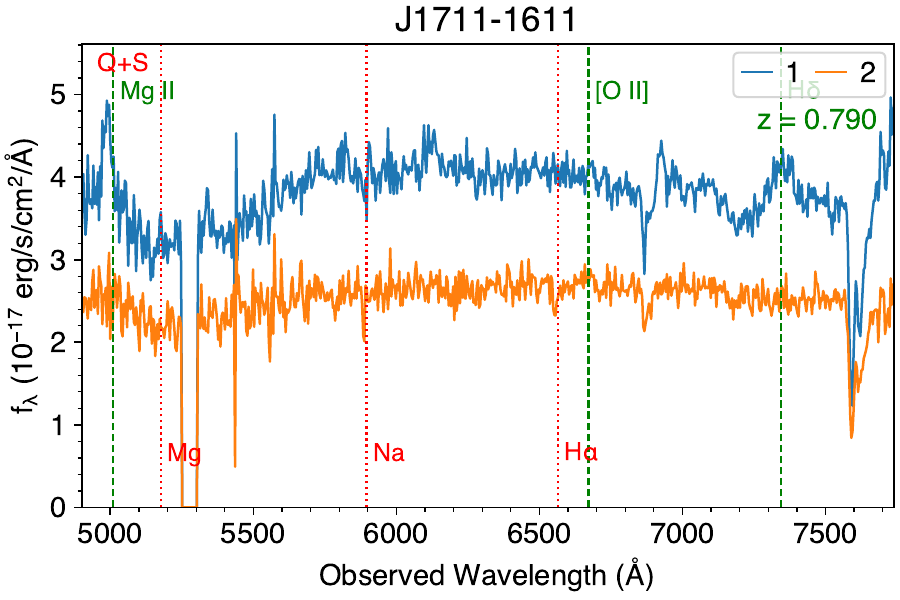}
    \includegraphics[width=0.32\textwidth]{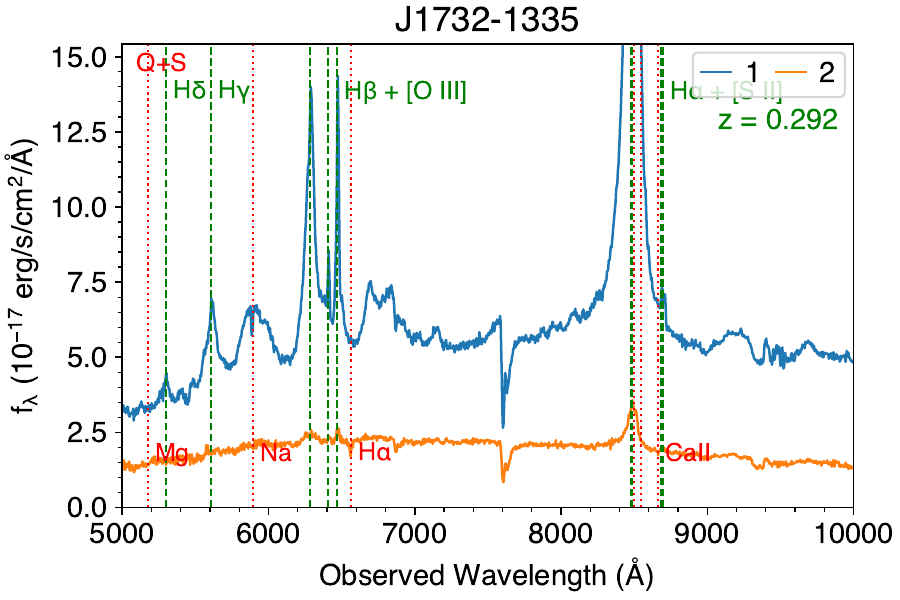}
    \includegraphics[width=0.32\textwidth]{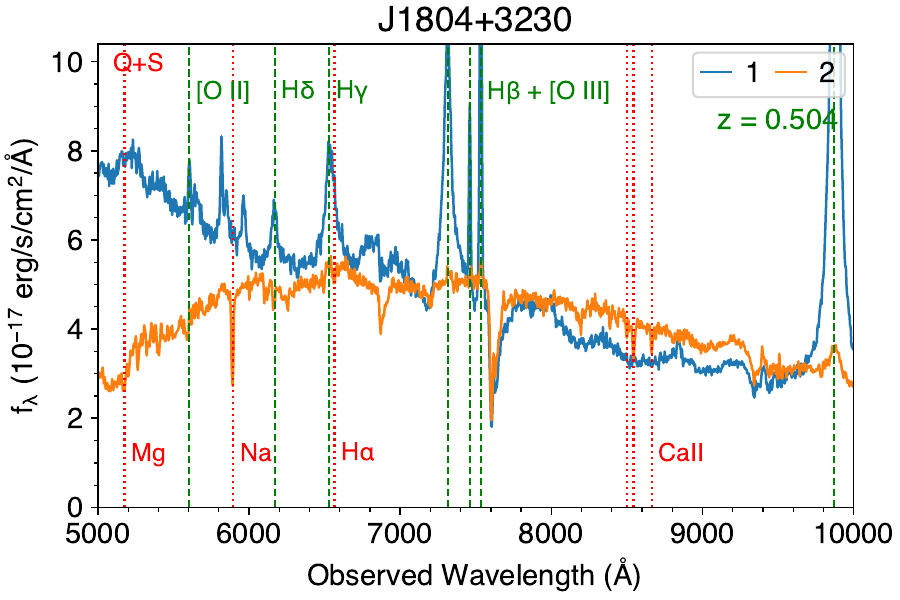}
    \includegraphics[width=0.32\textwidth]{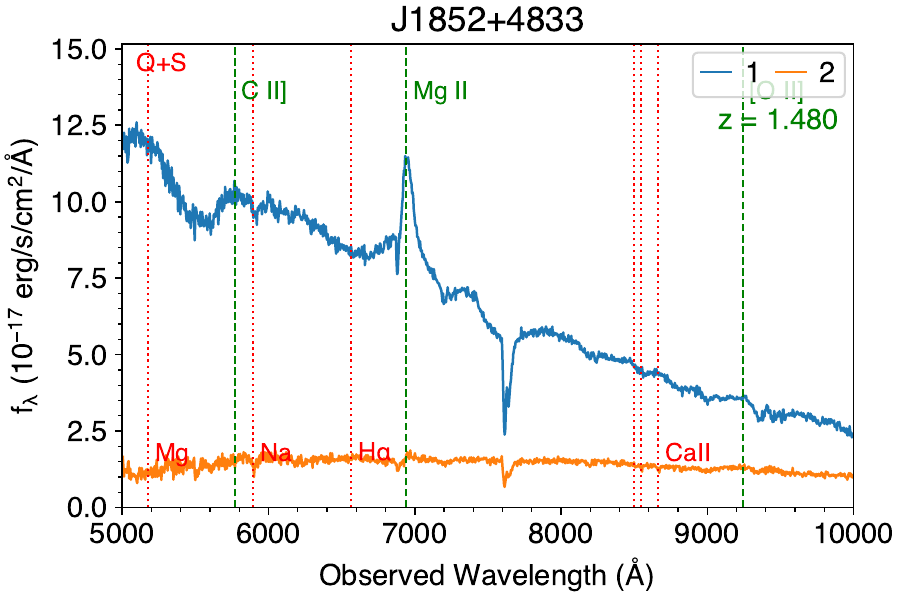}
    \includegraphics[width=0.32\textwidth]{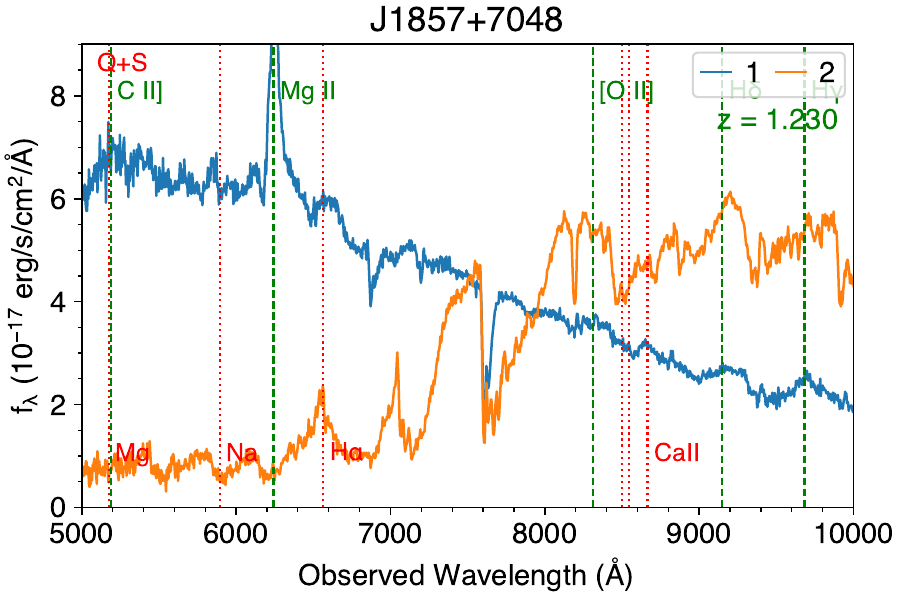}
    \includegraphics[width=0.32\textwidth]{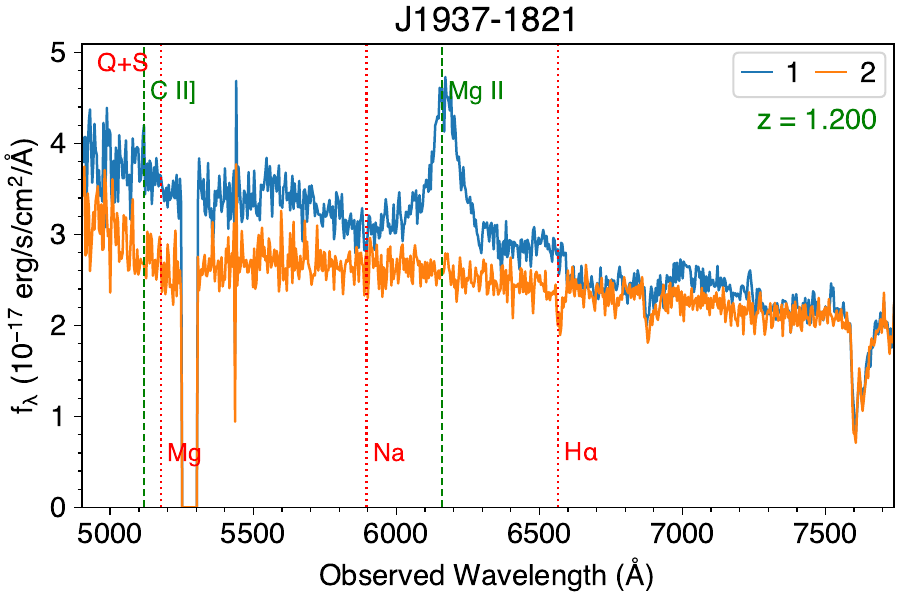}
    \includegraphics[width=0.32\textwidth]{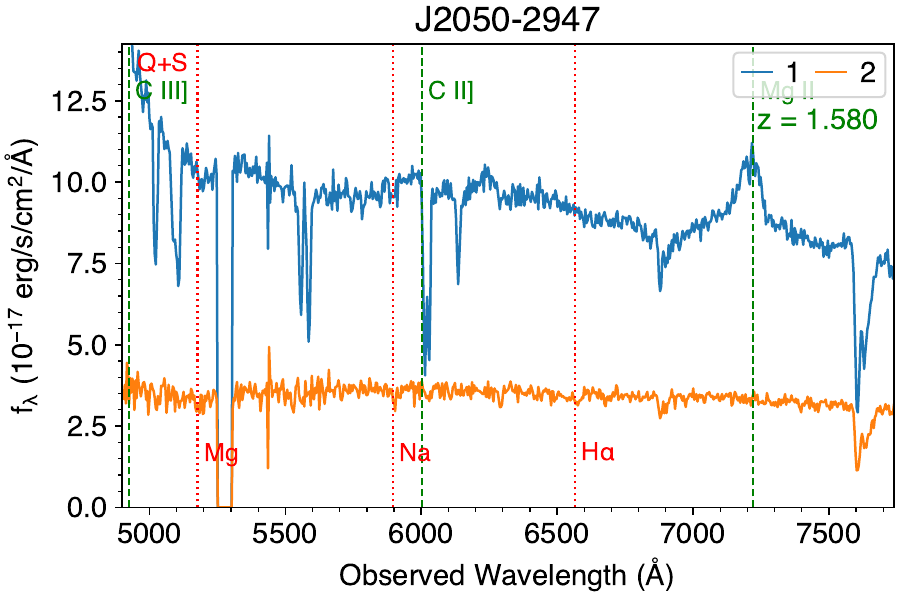}
    \includegraphics[width=0.32\textwidth]{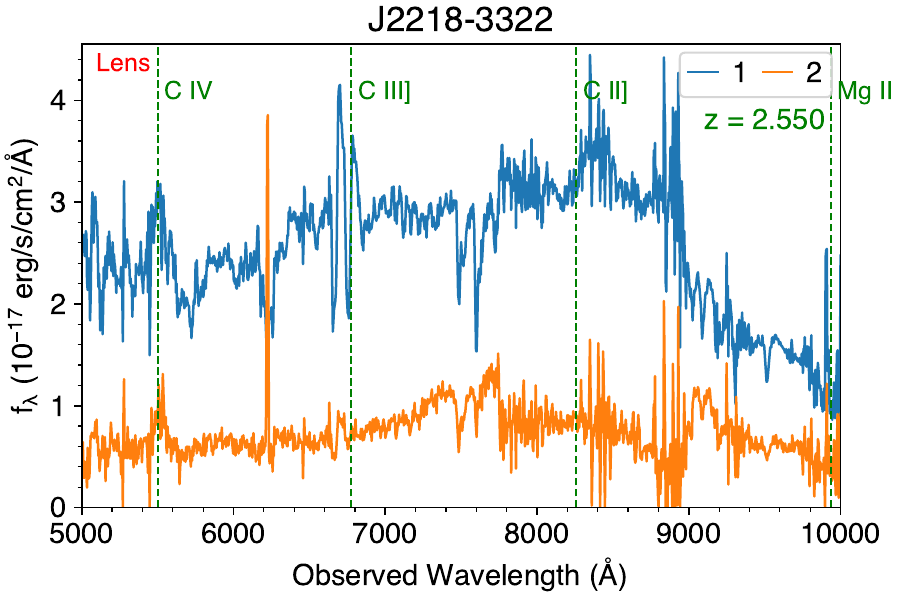}
    
    \caption{Gemini GMOS optical spectra of 16 targets. Spatially resolved spectra for both sources are shown in solid blue and dashed orange lines. Green vertical dashed lines mark the quasar's emission lines at the systemic redshift, displayed in the top-right corner. Red vertical dotted lines indicate stellar absorption features for star superposition.  Final classifications based on both Gemini and/or HST spectra are shown in the top-left corner (Q: quasar, S: star, U: unknown, Dual: dual quasar, Lens: lensed quasar).
    }
    \label{fig:spectra_gmos}
\end{figure*}

\autoref{fig:spectra_gmos} shows the Gemini/GMOS spectra of 16 targets. Using the two-Gaussian component fitting method described in \autoref{sec:data-gemini}, we successfully decompose the two sources for most of our targets. We also examine the 1D spatial profiles of the wavelength-averaged spectra to ensure that the two nuclei are well modeled by Gaussian components and that their flux ratios are consistent with those observed in HST optical images. Some targets, including J1649+0812 and J1732-1335, are affected by contamination from the adjacent bright nucleus, which we discuss in the individual classifications in \autoref{sec:classification}. Thanks to the higher signal-to-noise ratios compared to the STIS spectra, we can identify not only the broad emission lines from quasars but also the narrow emission lines and absorption lines from star interlopers. For the targets observed in program GS-2022A-Q-148, the spectra are truncated at 7700\AA\ due to poor response in one of the CCDs.

\subsection{Classifications} \label{sec:classification}
Using the spectra from Gemini and HST, we perform the following steps to classify each target:
\begin{enumerate}
    \item Identify quasars based on their broad emission lines and measure their redshifts.
    \item Identify stars by detecting absorption lines, such as NaD, if present.
    \item For featureless spectra without detected emission or absorption lines, identify quasars using deep radio imaging from \citet{Gross2024}. 
\end{enumerate}

Based on the spectra and auxiliary multi-wavelength images, we identify 7 dual/lensed quasars, 10 quasar-star superpositions, and 1 binary star. For the remaining 9 targets, 2 are likely dual/lensed quasars. We are unable to conclusively determine the nature of the other 7 targets but provide our best estimates based on the images and spectra. Each group of targets is discussed in the following subsections.

\subsubsection{Binary stars and quasar-star superpositions}

We detect at least one stellar absorption line (e.g., Na \textsc{i}, H$\alpha$, Ca \textsc{ii}) from nearby foreground stars in the following 11 targets:
\begin{itemize}
    \item J0246+6922: A binary M-type star. Both sources exhibit broad, strong Ca \textsc{ii} and TiO absorption lines, and extremely red colors in the HST spectra.
    
    \item J0904+3332: $z=1.106$ quasar + star. Source 1 shows broad Mg \textsc{ii} emission lines at $z=1.106$ in both HST and Gemini spectra. Source 2 shows Mg \textsc{i}, Na \textsc{i} and Ca \textsc{ii} absorption lines in the Gemini spectrum.
    
    \item J1314$-$4912: $z=2.295$ quasar + star. Source 1 shows broad C \textsc{iv}, C \textsc{iii}] and Mg \textsc{ii} emission lines at $z=2.295$, while Source 2 shows Mg \textsc{i}, Na \textsc{i} and Ca \textsc{ii} absorption lines in the Gemini spectra.
    
    \item J1649+0812: $z=1.391$ quasar + star. Source 1 shows broad Mg \textsc{ii} emission lines at $z = 1.391$ in both the HST and Gemini spectra. Source 2 exhibits H$\alpha$ and Ca \textsc{ii} absorption lines in the Gemini spectrum. The broad Mg \textsc{ii} and narrow [O \textsc{ii}] emission lines seen in Source 2 in the Gemini spectra are due to contamination from Source 1. High angular resolution VLT/MUSE spectra (private communication) confirm the presence of the same absorption features in Source 2, without any accompanying emission lines.
    
    \item J1711-1611: $z=0.790$ quasar + star. Source 1 shows broad H$\beta$ and H$\gamma$ emission lines in the HST spectrum, and broad Mg \textsc{ii} emission lines in the Gemini spectrum at $z=0.790$. Source 2 shows Mg \textsc{i}, Na \textsc{i} and H$\alpha$ absorption lines in the Gemini spectrum.
    
    \item J1732$-$1335: $z=0.292$ quasar + star. Source 1 shows broad H$\alpha$, H$\beta$ and H$\gamma$ emission lines at $z=0.292$ and Source 2 shows H$\alpha$ absorption lines in the Gemini spectra. The emission lines (e.g., H$\alpha$ and H$\beta$) seen in Source 2 are due to contamination from Source 1.
    
    \item J1804+3230: $z=0.504$ quasar + star. Source 1 shows broad H$\alpha$, H$\beta$ and H$\gamma$ emission lines at $z=0.504$ and Source 2 shows Na \textsc{i} and Ca \textsc{ii} absorption lines in the Gemini spectra.
    
    \item J1852+4833: $z=1.480$ quasar + star. Source 1 shows broad Mg \textsc{ii} emission lines at $z=1.480$ in both HST and Gemini spectra. Source 2 shows Na \textsc{i} and Ca \textsc{ii} absorption lines in the Gemini spectrum.
    
    \item J1857+7048: $z=1.230$ quasar + star. Source 1 shows broad Mg \textsc{ii} emission lines at $z=1.230$ and Source 2 shows broad, strong Ca \textsc{ii} and TiO absorption lines in the Gemini spectra.
    
    \item J1937-1821: $z=1.200$ quasar + star. Source 1 shows broad Mg \textsc{ii} emission lines at $z=1.200$ in both HST and Gemini spectra. Source 2 shows H$\alpha$ absorption line in the Gemini spectrum.
    \item J2050-2947: $z=1.580$ quasar + star. Source 1 shows broad Mg \textsc{ii} emission lines at $z=1.580$ in both HST and Gemini spectra. Source 2 shows Mg \textsc{i} and Na \textsc{i} absorption lines in the Gemini spectrum.
\end{itemize}

\subsubsection{Dual/lensed quasars}

For the following targets, we detect broad emission lines in both sources. Distinguishing between dual quasars and lensed quasars is challenging because the spectra of both sources in most of our dual/lensed quasar candidates are very similar. For example, J0749+2255 and J0823+2418 were confirmed as a dual quasar and a lensed quasar, respectively, based on additional multi-wavelength observations \citep{ChenYC2023a, Gross2023}. However, in the absence of definitive evidence from other observations or the literature, we do not distinguish between dual and lensed quasars in this paper. The classifications are listed below.

\begin{figure*}
  \centering
    \includegraphics[width=0.32\textwidth]{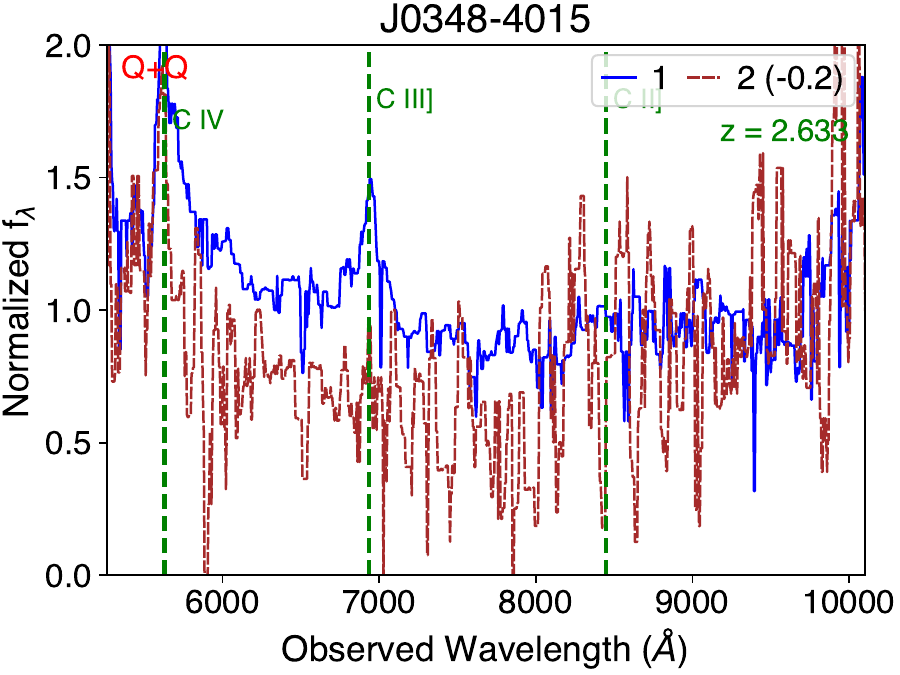}
    \includegraphics[width=0.32\textwidth]{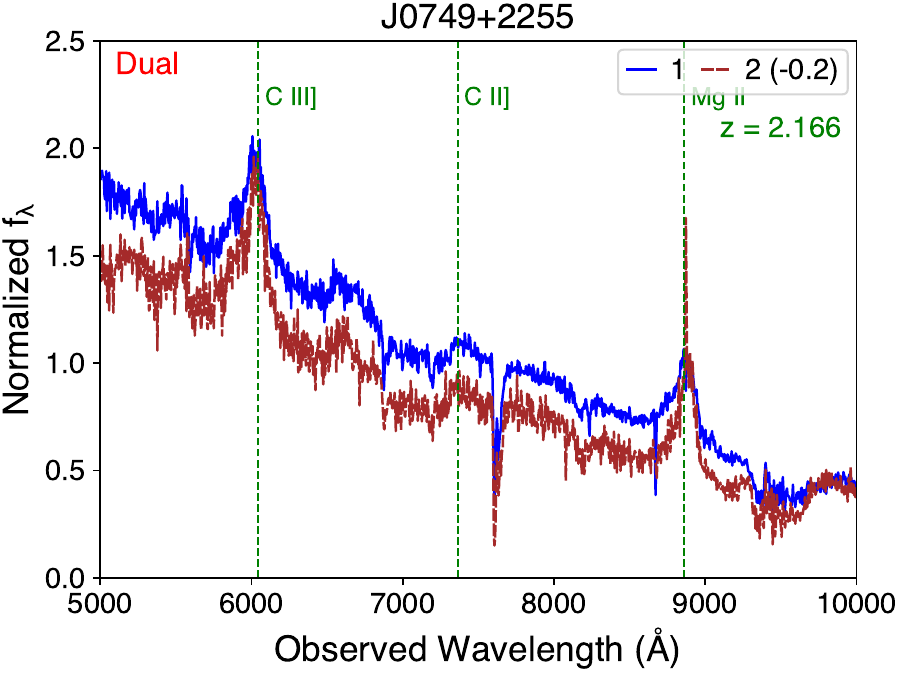}
    \includegraphics[width=0.32\textwidth]{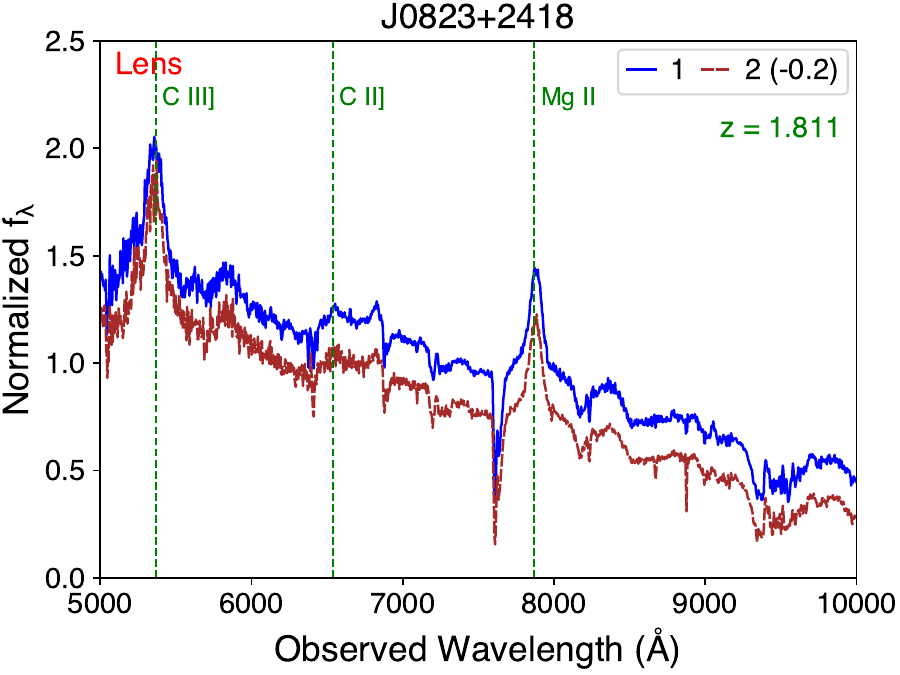}
    \includegraphics[width=0.32\textwidth]{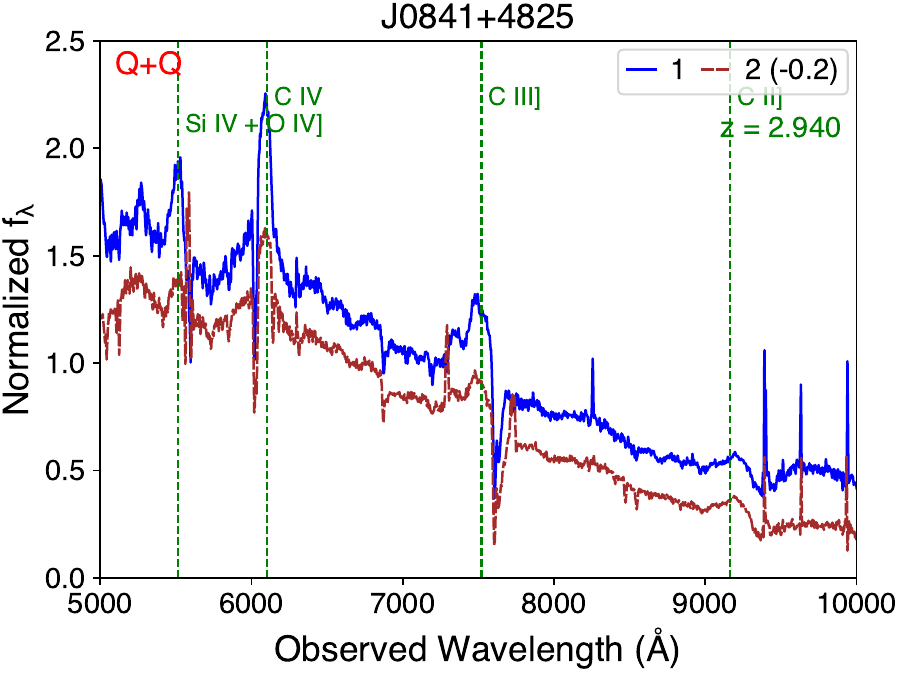}
    \includegraphics[width=0.32\textwidth]{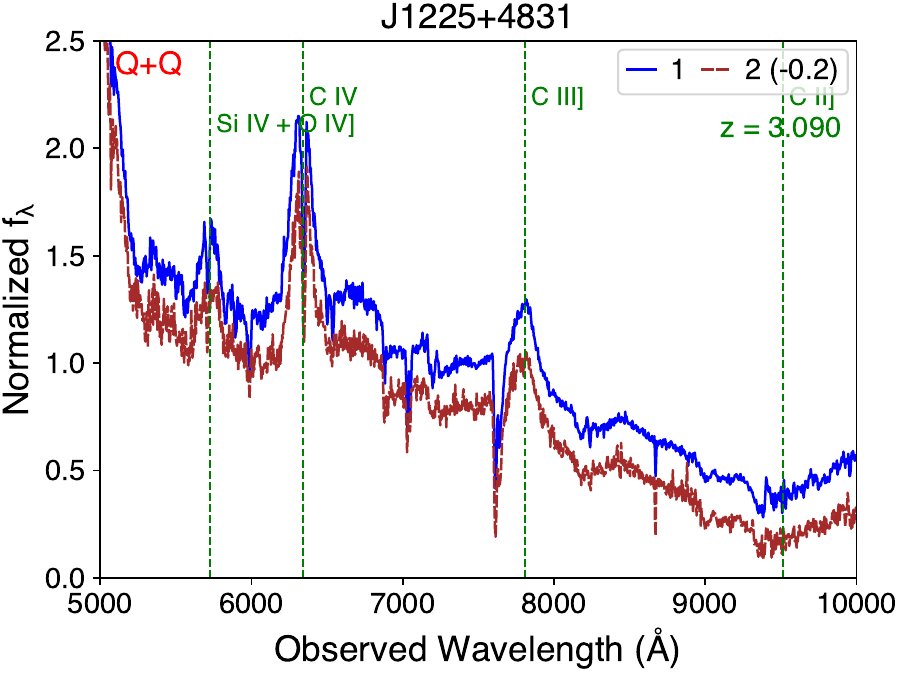}
    \includegraphics[width=0.32\textwidth]{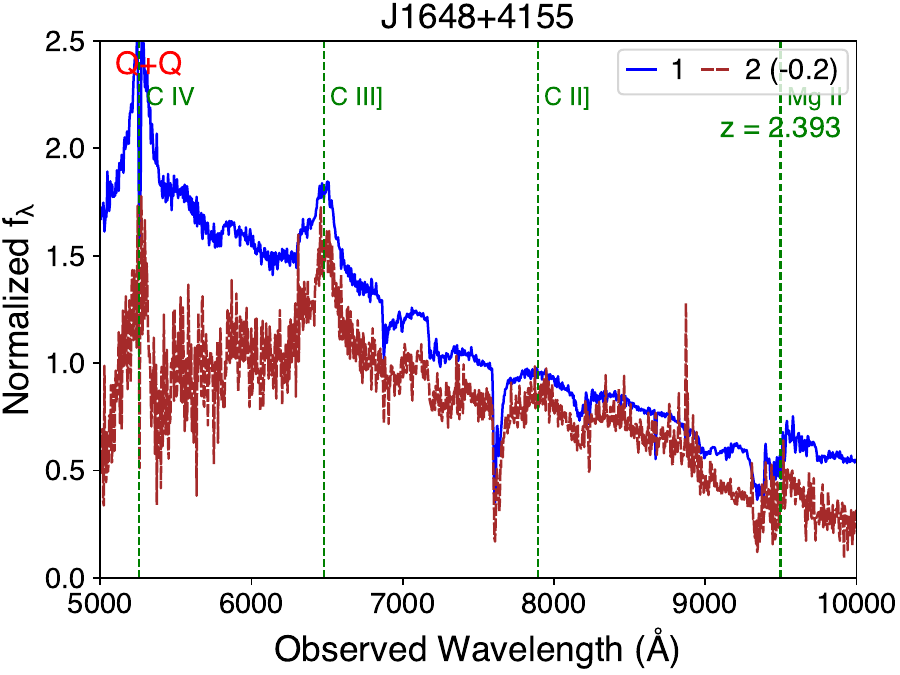}
    \includegraphics[width=0.32\textwidth]{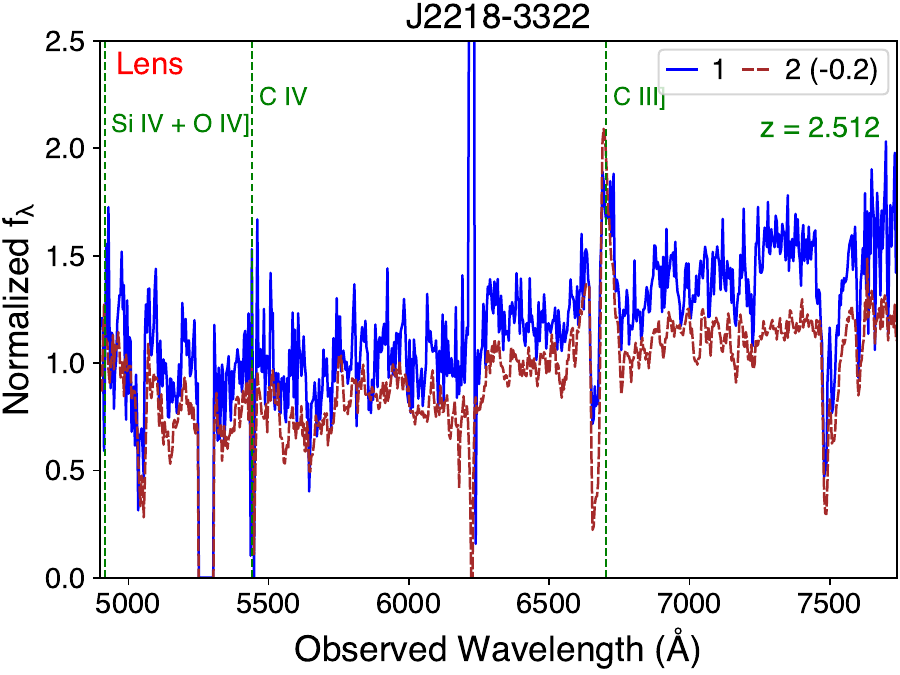}

    \caption{ Median-normalized spectra of the 7 dual/lensed quasars, with the second source spectra shifted by -0.2 to prevent overlap. Spatially resolved spectra for both sources are represented by solid blue and dashed brown lines. Quasar emission lines at the corresponding redshift (indicated in the top-right corner) are marked by green vertical lines.
    }
    \label{fig:spectra_all_dual_lens}
\end{figure*}

\begin{itemize}
    \item J0348$-$4015: dual/lensed quasar at $z=2.633$. Both sources show broad C \textsc{iv} emission lines at $z=2.633$ in the HST spectra. The low signal-to-noise ratio in the HST spectra prevents us from conducting further analysis. 
    
    \item J0749+2255: dual quasar at $z=2.166$ Both sources show broad C \textsc{iii}] and Mg \textsc{ii} emission lines at $z=2.166$ in the Gemini spectra. \citet{ChenYC2023a} detected the host galaxy and tidal tails through multiwavelength observations, confirming the system as a dual quasar.
    
    \item J0823+2418: lensed quasar at $z=1.811$. Both sources display broad C \textsc{iii}] and Mg \textsc{ii} emission lines at $z=1.811$ in both the Gemini and HST spectra. \citet{Gross2023} identified the lensing galaxy in the HST near-infrared image, and the relative positions of the components are consistent with the lensing model.
    
    \item J0841+4825: dual/lensed quasar at $z=2.940$. Both sources exhibit broad C \textsc{iii}] and C \textsc{iv} emission lines at $z=2.940$ in both the Gemini and HST spectra. \citet{Mannucci2022} classified the system as a dual quasar based on the presence of weak C \textsc{iii}]1909 and N\textsc{iii}]1750 lines detected only in the secondary source in the HST spectrum. However, we do not observe these features in the higher signal-to-noise Gemini spectrum. Given that the two sources display very similar spectral features in the Gemini data, we consider both the dual and lensed quasar scenarios to be plausible.
    
    \item J1225+4831: dual/lensed quasar at $z=3.090$. Both sources exhibit broad C \textsc{iii}] and C \textsc{iv} emission lines at $z=3.090$ in the Gemini spectra.
    
    \item J1648+4155: dual/lensed quasar at $z=2.393$. Both sources exhibit broad H$\alpha$ emission lines at $z=2.393$ in both the Gemini and HST spectra. The broad H$\alpha$ emission in Source 2 is unlikely due to spectral contamination from Source 1 because it also presents in the HST spectrum.
    
    \item J2218$-$3322: quadruply lensed broad absorption line (BAL) quasar at $z=2.550$. HST optical imaging reveals four background quasar images arranged around a foreground lens galaxy \citep{ChenYC2022}. Our Gemini spectra confirm the background quasar's redshift at $z=2.550$ with the C \textsc{iv} emission line and show that it is a BAL quasar.
\end{itemize}

\autoref{fig:spectra_all_dual_lens} shows the median-normalized spectra of the 7 dual/lensed quasars, highlighting the similarities and differences between the two spectra for each system.

\subsubsection{Unknown cases}

For the 9 remaining targets, we either detect only one broad emission line or are unable to identify any emission or absorption lines. Most of the sources only have low signal-to-noise HST spectra, which prevent us to identify any absorption lines. We also examine radio observations for some targets \citep{Gross2024} to identify potential obscured quasars.

\begin{itemize}
    \item J0241+7801: Both sources are likely quasars with unknown redshifts. Source 1 shows a possible emission line at 8500 \AA in the HST spectrum, while Source 2 has a radio detection of $>$2mJy \citep{Gross2024}.  
    \item J0536+5038: $z=1.855$ quasar + unknown source. Source 1 shows broad Mg \textsc{ii} and C \textsc{iii}] emission lines at $z=1.855$, while Source 2 shows a featureless flat spectrum in the HST spectra.
    \item J0748+3146: $z=1.407$ quasar + unknown source. Source 1 exhibits broad Mg \textsc{ii} emission lines at $z=1.407$, while Source 2 shows no significant spectral features in the HST spectra.
    \item J0753+4247: $z=1.523$ quasar + unknown source. Source 1 exhibits broad Mg \textsc{ii} and C \textsc{ii}] emission lines at $z=1.523$, while Source 2 shows no significant spectral features in the HST spectra.
    \item J1327+1036: $z=1.904$ quasar + a likely quasar. Source 1 exhibits broad Mg \textsc{ii} and C \textsc{iii}] emission lines at $z=1.904$, while Source 2 shows no significant spectral features in the HST spectra. Both sources are detected in the radio at the mJy level, likely indicating quasar activity \citep{Gross2024}. 
    \item J1613-2644: Unknown case. No emission lines are detected in both sources in the HST spectra.
    \item J1755+4229: $z=1.908$ quasar + unknown source. Source 1 exhibits broad Mg \textsc{ii} and C \textsc{iii}] emission lines at $z=1.908$, while Source 2 shows no significant spectral features in the HST spectra.
    \item J2048+6258: $z=2.420$ quasar + unknown source. Source 1 exhibits broad Mg \textsc{ii}, C \textsc{ii}] and C \textsc{iii}] emission lines at $z=2.420$, while Source 2 shows no significant spectral features in the HST spectra.
    \item J2212-0026: $z=1.975$ quasar + unknown source. Source 1 exhibits broad Mg \textsc{ii} and C \textsc{iii}] emission lines at $z=1.975$, while Source 2 shows no significant spectral features in the HST spectra.
\end{itemize}

\begin{deluxetable*}{ccccccc}
\tablecaption{Properties and classifications of the 27 targets\label{tab:target}}
\tablehead{\colhead{Abbreviated Name} & \colhead{R.A.} & \colhead{Decl.} & \colhead{Sep.} & \colhead{Target Cat.} & \colhead{Color Selection} & \colhead{Spectroscopic Classification} \\
\colhead{(J2000)} & \colhead{(deg)} & \colhead{(deg)} & \colhead{(\arcsec)} & & }
 \colnumbers
 \startdata
WISE J0241+7801 & 40.395481 & 78.018624 & 0.58 & 2 & quasar+star & likely quasar+quasar\\
WISE J0246+6922 & 41.619305 &
69.376122 & 0.49 & 5 & quasar+star & \textbf{binary star} \\
WISE J0348$-$4015 & 57.119449 & 
-40.253662 & 0.50 & 3 & dual/lensed quasar & \textbf{dual/lensed quasar(z=2.633)} \\
WISE J0536+5038 & 84.084298 &
50.6406250 & 0.32 & 2 & dual/lensed quasar & quasar(z=1.855) + unknown\\
SDSS J0748+3146 & 117.002307 & 
31.779856 & 0.53 & 4 & dual/lensed quasar & quasar(z=1.407) + unknown \\
SDSS J0749+2255 & 117.345694 & 22.919936 & 0.46 & 1 & dual/lensed quasar & \textbf{dual quasar(z=2.166)}\\
SDSS J0753+4247 & 118.460738 &
42.795529 & 0.33 & 1 & quasar+star & quasar(z=1.523) + unknown \\
SDSS J0823+2418 & 125.921169 & 
24.301570 & 0.64 & 4 & dual/lensed quasar & \textbf{lensed quasar(z=1.811)}\\
SDSS J0841+4825 & 130.374046 & 
48.430135 & 0.46 & 1 & dual/lensed quasar & \textbf{dual/lensed quasar(z=2.940)}\\
SDSS J0904+3332 & 136.036118 &
33.534796 & 0.30 & 1 & quasar+star & \textbf{quasar(z=1.106) + star}\\
SDSS J1225+4831 & 186.327776 & 
48.521147 & 0.90 & 4\tablenotemark{a} &  \nodata\tablenotemark{a} & \textbf{dual/lensed quasar(z=3.090)}\\
WISE J1314$-$4912 & 198.566742 & 
-49.205070 & 0.52 & 6 & quasar+stars &\textbf{quasar(z=2.295) + star}\\
SDSS J1327+1036 & 201.966887 & 
10.607564 & 0.75 & 1\tablenotemark{a} &  \nodata\tablenotemark{a} & quasar(z=1.904)+ likely quasar\\
WISE J1613$-$2644 & 243.456316 &
-26.742377 & 0.28 & 2 & quasar+star & unknown \\
SDSS J1648+4155 & 252.075328 &
41.930608 & 0.44 & 1 & dual/lensed quasar &\textbf{dual/lensed quasar(z=2.393)}\\
WISE J1649+0812 & 252.422063 &
8.209311 & 0.59 & 5 & dual/lensed quasar &\textbf{quasar(z=1.391) + star}\\
WISE J1711$-$1611 & 257.916584 &
-16.196649 & 0.67 & 5 & dual/lensed quasar & \textbf{quasar(z=0.790) + star}\\ 
WISE J1732$-$1335 & 263.095334 &
-13.593137 & 0.72 & 2 & dual/lensed quasar+star &\textbf{quasar(z=0.292) + star}\\
WISE J1755+4229 & 268.929920 & 
42.490069 & 0.59 & 5 & quasar+star & quasar(z=1.908) + unknown \\
WISE J1804+3230 & 271.039815 &
32.508215 & 0.68 & 5 & quasar+star &\textbf{quasar(z=0.504) + star}\\
WISE J1852+4833 & 283.108771 &
48.554173 & 0.62 & 2 & quasar+star &\textbf{quasar(z=1.480) + star}\\
WISE J1857+7048 & 284.369408 &
70.803154 & 0.61 & 5 & quasar+star & \textbf{quasar(z=1.230) + star}\\
WISE J1937$-$1821 & 294.328398 &
-18.358950 & 0.62 & 2 & dual/lensed quasar & \textbf{quasar(z=1.200) + star}\\ 
WISE J2048+6258 & 312.200003 &
62.98286 & 0.62 & 5 & dual/lensed quasar & quasar(z=2.420) + unknown \\
WISE J2050$-$2947 & 312.500061 & 
-29.789362 & 0.65 & 5 & dual/lensed quasar &\textbf{quasar(z=1.580) + star} \\
SDSS J2122$-$0026 & 320.679218 &
-0.448282 & 0.52 & 4 & dual/lensed quasar & quasar(z=1.975) + unknown 
\\
WISE J2218$-$3322 & 334.707755 & 
-33.378782  & 0.49 & 5 & quad lens & \textbf{quad lens(z=2.512)} \\
\enddata
 \tablecomments{Column 1: Target name. Column 2: Right Ascension. Column 3: Declination. Column 4: Pair separation in arcsecond. Column 5: Target category from \citet{ChenYC2022} (1: SDSS Gaia-unresolved, 2: WISE+PS1 Gaia-unresolved, 3: WISE-only Gaia-unresolved, 4: SDSS Gaia-resolved, 5: WISE+PS1 Gaia-resolved, 6: WISE-only Gaia-resolved). Column 6: Photometric classification using HST F475W and F814W images from \citet{ChenYC2022}. Column 8: Spectroscopic classification based on the optical spectra in this paper. Entries in boldface indicate confirmed classifications.}
 \tablenotetext{a}{J1225+4831 and J1327+1036 were not observed in HST snapshot imaging program \citep{ChenYC2022}, but they were discovered using the same VODKA technique.}
\end{deluxetable*}

\section{Discussion} \label{sec:discussion}

\subsection{Quasar-star-superposition and dual/lensed quasar fractions}

Based on the 2-band color cut using HST F475W and F814W images, \citet{ChenYC2022} estimated a quasar-star superposition rate of 30\%$\pm$10\% for the targets in the VODKA project. However, this simple 2-band color cut method primarily differentiates late-type stars from typical unobscured quasars, potentially misclassifying early-type stars and obscured quasars. With our spectroscopic follow-up observations, we can now provide a more accurate assessment of the fractions of star superpositions and dual/lensed quasars.

Among the 45 targets discovered in the VODKA imaging project \citep{ChenYC2022} and additional targets from various subsequent campaigns, we observed 25 with our spectroscopic data. For these observed targets, we calculated a star superposition rate of 11/27 = 41\%. This is likely a lower limit because we selectively followed up on targets with blue and similar colors, and additional quasar-star superpositions may be present among the uncertain cases. Assuming that targets with unknown companions are all quasar-star superpositions, the upper limit of the quasar-star superposition rate is 18/27 = 67\%. This 41-67\% superposition rate exceeds the estimated rate of 30\%$\pm$10\% from the two-band color cut \citep{ChenYC2022}. The higher rate is likely due to the exclusion of stars (e.g., F and G stars) with similar optical colors to quasars in the tow-band color cut. For dual/lensed quasars, counting only those confirmed spectroscopically from spatially resolved spectra, we obtained a fraction of 7/27 = 26\%. This fraction is a lower limit, as additional dual/lensed quasars might not have been discovered due to low signal-to-noise spectra or because they were not observed due to redder colors.

We can evaluate whether additional selection criteria could yield a cleaner sample. For instance, if we restrict our candidates to those identified as spectroscopically confirmed quasars, and apply the same reasoning as before, we find that the quasar-star superposition rate ranges from 1/10 = 10\% to 4/10 = 40\%. Additionally, the lower limit for the fraction of dual/lensed quasars increases to 5/10 = 50\%.

We can examine whether extra selection criteria can provide a cleaner sample. For example, if we only select candidates from spectroscopically confirmed quasars and following the same argument in the previous paragraph, we can obtain the quasar-star superposition rate between 1/10 = 10\% and 4/10 = 40\%, and the lower limit of dual/lensed quasar fraction is 5/10 = 50\%. 
Similar calculations can be performed by considering only the targets classified as dual/lensed quasar candidates using HST F475W and F814W images \citep{ChenYC2022}. The quasar-star superposition and dual/lensed quasar fractions for various samples are listed in \autoref{tab:fraction}. In summary, with existing unresolved spectra from SDSS and the spatially resolved two-band color cut, the fraction of dual/lensed quasars can be pushed to higher than 67\%.

\begin{deluxetable}{lccc}
\tablecaption{Fraction of quasar-star superpostions and dual/lensed quasars \label{tab:fraction}}
\tablehead{\colhead{Sample} &  \colhead{\#}& \colhead{Dual/lens} & \colhead{Quasar-star}  \\
\colhead{(1)} & \colhead{(2)} & \colhead{(3)} & \colhead{(4)} }
    \startdata
       1. Whole follow-up sample &  25 & $\gtrsim$26\% & $\sim$41-67\% \\
        2. SDSS quasars only &  10 &$\gtrsim$50\% & $\sim$10-40\%\\
    3. F475W-F814W color cut &  15& $\gtrsim$40\% &  $\sim$33-60\%\\
    4. Criteria 2 and 3 & 6 &  $\gtrsim$67\% &  $\sim$0-33\% \\
        \hline
    \enddata
    \tablecomments{Column 1: Sample selection. Column 2: Number of targets. Column 3: Dual/lensed quasar fraction 4: Quasar-star superpostion fraction. F475W-F814W color cut is mainly based on the photometric classification in \citet{ChenYC2022}.}
\end{deluxetable}

\subsection{J1649+0812: A cautionary tale of sources with velocity offsets}

\begin{figure}
  \centering
    \includegraphics[width=0.8\columnwidth]{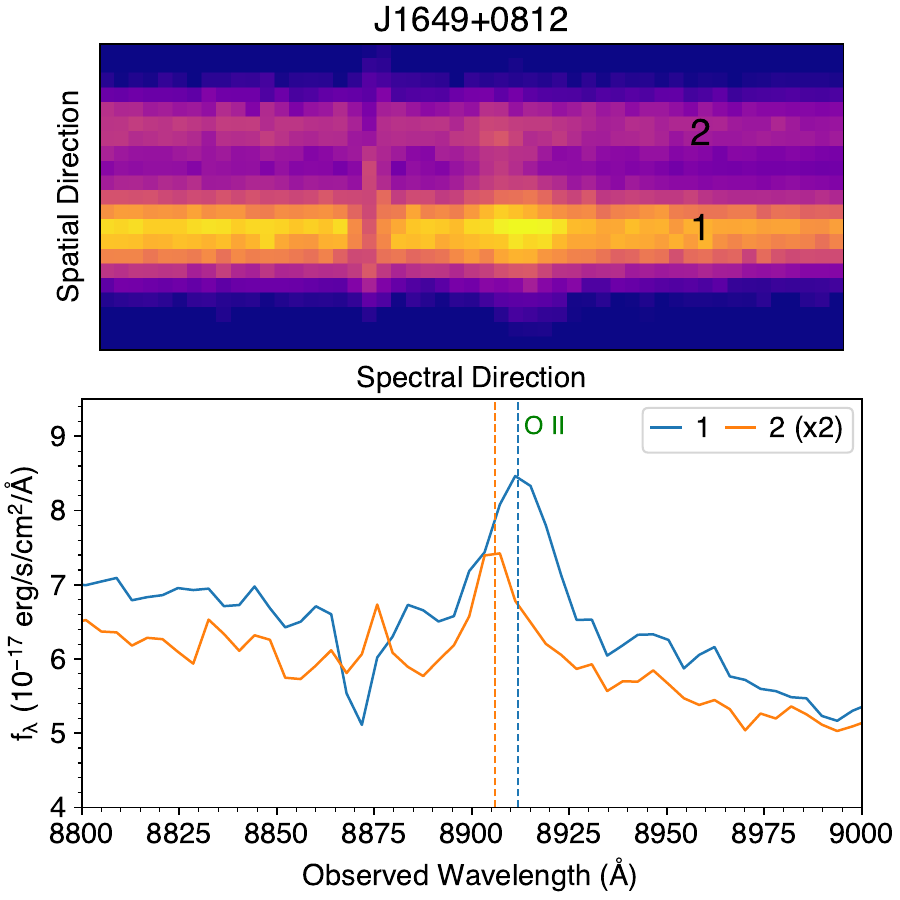}
    
    \caption{Top: Marginally resolved 2D spectrum of J1649+0812, showing the traces of two sources and the [O \textsc{ii}] emission. Bottom: 
    Extracted 1D spectra of J1649+0812, showing a possible velocity offset of $\sim$200 km/s between the [O \textsc{ii}] emission lines. However, given that Source 2 is a star, this velocity offset is likely due to internal orbital motion within a single host galaxy. The [O \textsc{ii}] emission line observed in Source 2 likely originates from the extended host galaxy of Source 1, rather than from Source 2 itself.
    }
    \label{fig:spectra_J1649}
\end{figure}

J1649+0812 was initially classified as a dual quasar when we submitted the draft, based on the detection of a broad Mg \textsc{ii} emission line in Source 2 and a possible velocity offset of 183$\pm$76 km/s between the two sources in the narrow [O \textsc{ii}] emission line (\autoref{fig:spectra_J1649}). However, further analysis of the Gemini spectrum reveals that Source 2 is a star, as indicated by the presence of H$\alpha$ and Ca \textsc{ii} absorption lines.
The AO-assisted VLT/MUSE data with higher angular resolution (FWHM $\sim$ 0\farcs1) and improved signal-to-noise ratio (private communication) also show that the weak broad Mg \textsc{ii} feature is likely an artifact of the marginally resolved Gemini spectra (FWHM $\sim$ 0\farcs5), and is not detected in the resolved HST spectrum (FWHM $\sim$ 0\farcs1). The $\sim$200 km/s velocity offset in [O \textsc{ii}] likely arises from internal orbital motion within a single host galaxy. The [O \textsc{ii}] emission line observed in Source 2 likely originates from the extended host galaxy of Source 1, rather than from Source 2 itself. We caution that small velocity offsets ($\lesssim$300 km/s) in narrow emission lines should not be used as standalone evidence for identifying dual quasars, particularly in marginally resolved spectra.

\subsection{Implication for future spectroscopic follow-ups}

Based on the results of this study, we outline the following implications for future spectroscopic follow-up to confirm the nature of dual quasar candidates:
\begin{itemize}

    \item Given the high star-superposition rate (37–63\%), we suggest performing detailed spectral decomposition for targets with existing spectra (e.g., SDSS) before conducting follow-up observations. \citet{Shen2023} successfully used spectral principal component analysis to identify potential star superpositions, revealing star companions of various types. Applying similar spectral decomposition to existing spectra can help reduce the star-superposition rate and better allocate limited telescope time to higher-confidence dual/lensed quasar candidates.
    
    \item A decent signal-to-noise ratio for the continuum is essential to detect absorption lines such as Na \textsc{i}, \ha, and Mg \textsc{i}, particularly for F and G type stars. We recommend an signal-to-noise ratio of $\gtrsim$ 20 per spectral element. In our data, the Gemini spectra achieve an signal-to-noise ratio of at least $\sim$20 per spectral element, whereas the HST/STIS spectra range from 1 to 10 per spectral element.

    \item  High spatial resolution is essential to avoid spectral contamination from the primary source to the secondary source. Some of our targets suffer from blending in the Gemini spectra, which have angular resolutions comparable to the separations between sources. We recommend using spatial resolution at least twice as high (i.e., FWHM no more than half the source separation) to reliably disentangle the spectra of close pairs.
    
    \item High spectral resolution may be necessary to reliably identify velocity offsets in dual quasars. However, small velocity offsets should not be used as standalone evidence for dual quasar identification, especially in marginally resolved spectra. We recommend a spectral resolution of $R \gtrsim 1000$ (corresponding to $\sim$300 km/s), which is sufficient to resolve offsets larger than typical galaxy rotation velocities.

\end{itemize}




\section{Conclusion} \label{sec:conclusion}

In this paper, we present extensive spectroscopic follow-up observations of 27 dual quasar candidates identified by the VODKA project. Using data from Gemini/GMOS and HST/STIS, we classify and assess the nature of these systems. We identify 11 cases as star–quasar superpositions and 7 systems as either dual or lensed quasars. For the remaining 9 targets, a conclusive classification is not possible, although 2 are likely dual quasars based on supporting radio data. Among the 7 dual/lensed quasars, 3 had been previously confirmed in the literature. Additionally, we determine the redshift of the background quasar in a quadruply lensed system, for which the redshift was previously unknown.

Our follow-up observations yield a star-superposition rate of 41-67\% in the VODKA sample, which exceeds previous estimates. The fraction of dual/lensed quasars is at least 30\%. The higher quasar-star superposition rate is likely due to the lack of initial color selection or spectral decomposition for the initial VODKA sample. By leveraging unresolved SDSS spectra and spatially resolved two-band color cuts, we estimate that the dual/lensed quasar fraction could exceed 67\%. The ongoing HST snapshot program (Program ID: SNAP-17455; PI: Shen) incorporates various selection criteria to obtain a cleaner sample of dual/lensed quasar candidates. Further multi-wavelength observations and detailed analysis of narrow emission lines are necessary to better differentiate between dual and lensed quasars. Our study highlights the critical need for high-quality spectral data -- with a signal-to-noise ratio of $\gtrsim$20, spatial resolution at least twice finer than the source separation, and spectral resolution of $R \gtrsim 1000$ -- to effectively separate close sources, rule out stellar superpositions, and reliably identify dual quasars.

\begin{acknowledgments}


We thank the anonymous referee for giving constructive comments. We thank Michael Leveille, Alison Vick, Kristin Chiboucas, Hwihyun Kim, Trent Dupuy, Atsuko Nitta, Siyi Xu, and Rodolfo Angeloni for their help with our HST and Gemini observations. We thank Dr. Filippo Mannucci for his valuable and constructive comments.
%
This work is supported by NSF grant AST-2108162. Y.S. acknowledges partial support from NSF grant AST-2009947. Support for Program number HST-GO-16210 and HST-GO-16887 (PI: X. Liu) was provided by NASA through grants from the Space Telescope Science Institute, which is operated by the Association of Universities for Research in Astronomy, Incorporated, under NASA contract NAS5-26555. 
%
Based on observations made with the NASA/ESA Hubble Space Telescope, obtained from the Data Archive at the Space Telescope Science Institute, which is operated by the Association of Universities for Research in Astronomy, Inc., under NASA contract NAS 5-26555. These observations are associated with programs GO-16210 and GO-16887.


Based in part on observations obtained at the international Gemini Observatory (Program IDs GN-2020A-DD-106, GN-2022A-Q-139, and GS-2022A-Q-148; PI: X. Liu, and GN-2020A-Q-232; PI: Y.-C. Chen), a program of NSF's NOIRLab, which is managed by the Association of Universities for Research in Astronomy (AURA) under a cooperative agreement with the National Science Foundation. on behalf of the Gemini Observatory partnership: the National Science Foundation (United States), National Research Council (Canada), Agencia Nacional de Investigaci\'{o}n y Desarrollo (Chile), Ministerio de Ciencia, Tecnolog\'{i}a e Innovaci\'{o}n (Argentina), Minist\'{e}rio da Ci\^{e}ncia, Tecnologia, Inova\c{c}\~{o}es e Comunica\c{c}\~{o}es (Brazil), and Korea Astronomy and Space Science Institute (Republic of Korea). This work was enabled by observations made from the Gemini North telescope, located within the Maunakea Science Reserve and adjacent to the summit of Maunakea. We are grateful for the privilege of observing the Universe from a place that is unique in both its astronomical quality and its cultural significance.

\end{acknowledgments}

%

\vspace{5mm}
\facilities{HST(STIS), Gemini(GMOS)}


\software{
numpy \citep{numpy}, astropy \citep{Astropy2013,Astropy2018,Astropy2022}
}





\bibliography{ref}{}
\bibliographystyle{aasjournal}



\end{document}